\documentclass[useAMS,usenatbib]{mn2e}
\usepackage{graphicx,epsfig}

\def\deg{\hbox{$^\circ$}}

\title{The {\em Chandra} view of NGC\,1800 and the X-ray scaling properties of 
dwarf starbursts}

\author[J. Rasmussen, I. R. Stevens and T. J. Ponman]
{J.~Rasmussen$^1$\thanks{E-mail: jr@astro.ku.dk},
I.~R.~Stevens$^2$, T.~J. Ponman$^2$ \\
$^1$ Astronomical Observatory, University of Copenhagen, 
Juliane Maries Vej 30, DK-2100 Copenhagen \O, Denmark\\
$^2$ School of Physics and Astronomy, University of Birmingham,
Edgbaston, Birmingham B15 2TT, UK}

\date{}

\pagerange{\pageref{firstpage}--\pageref{lastpage}} \pubyear{2004}
\def\LaTeX{L\kern-.36em\raise.3ex\hbox{a}\kern-.15em
T\kern-.1667em\lower.7ex\hbox{E}\kern-.125emX}

\begin{document}
\label{firstpage}
\maketitle

\begin{abstract}
The superb spatial resolution of {\em Chandra} is utilized to study the
X-ray morphology of the dwarf starburst galaxy NGC\,1800 embedded in a small 
group of galaxies. Diffuse galactic emission is detected, extending several 
kpc above the galactic plane, with an overall morphology 
similar to the galactic winds seen in nearby X-ray bright starburst galaxies.
This makes NGC\,1800 the most distant dwarf starburst with a clear detection 
of diffuse X-ray emission. The diffuse X-ray luminosity of 
$1.3\pm 0.3\times 10^{38}$ erg s$^{-1}$ accounts for at least 60 per cent of 
the total soft X-ray output of the galaxy. 
A hot gas temperature of $kT=0.25$ keV and 
metallicity $Z\approx 0.05$Z$_{\odot}$ are derived, the latter in consistency
with results from optical spectroscopy of the interstellar medium.
Our failure to detect 
any hot gas associated with the embedding galaxy group translates into an 
upper limit to the group X-ray luminosity of $L_X<10^{41}$ erg s$^{-1}$. 
There is no convincing evidence that the outflowing wind of NGC\,1800 is 
currently interacting with any intragroup gas, and 
mechanical considerations indicate that the wind can escape the galaxy and 
its surrounding H{\sc i} halo,
eventually delivering energy and metals to the intragroup gas. 
Properties of NGC\,1800 are compared to those of other dwarf starburst 
galaxies, and a first detailed discussion of the X-ray scaling properties of 
this population of objects is given, set against the equivalent results 
obtained for normal starburst galaxies. Results indicate that dwarf starbursts
to a large degree behave as down-scaled versions of normal starburst
galaxies.
\end{abstract}

\begin{keywords}
ISM: jets and outflows -- galaxies: individual: NGC\,1800 -- galaxies: starburst -- galaxies: haloes -- X-rays: galaxies.
\end{keywords}

\section{Introduction}\label{sec,intro}

Starbursts, periods of short but intense star formation, and associated
galactic winds, are recognized as key elements
in the context of galaxy formation and evolution (e.g.\ \citealt*{heck1990}). 
Starburst winds also
constitute a viable candidate for the preheating mechanism 
(\citealt{kais1991}; \citealt{evra1991}),
thought to be at least partly responsible for the observed entropy level in 
the intergalactic medium of groups and clusters of galaxies, a level which is
poorly explained by models of structure formation invoking
gravitational interactions alone \citep*{ponm1999}.

Gas loss in galactic starburst winds is well-established, and can result
in a significant fraction of the interstellar medium being removed
(\citealt*{read1997}; \citealt{stri2000b}; \citealt{read2001}).

Whilst a rapidly moving cluster galaxy may be ram-pressure stripped
by its motion through the intracluster medium (ICM), much of the starburst 
activity
in denser environments is likely to take place in groups,
where velocities are too low for stripping to be a very effective process
\citep*{gaet1987}.
Under these circumstances, the confining
ICM might cause the wind to stall, and prevent gas escaping
from the galaxy (\citealt{babu1992}; \citealt{mura1999}).
This would have substantial implications both for galaxy evolution
(much reduced gas loss) and for the ICM (reduced metal enrichment
and energy injection). Bearing in mind that most galaxies are found in
groups (\citealt{tull1987}; \citealt{eke2004}), 
that a substantial fraction of star formation is bursty (\citealt{heck1998}
and references therein), and that
interactions in groups may actually trigger starburst activity 
(e.g.\ \citealt{zou1995}), results of this process are of great importance 
for our understanding of galaxy evolution and the evolutionary
relationship between group galaxies and their surroundings.

In the absence of a confining medium, simulations indicate that a superwind
will break out into the interstellar medium (ISM), driving a shell of 
swept-up material which
eventually fragments due to Rayleigh--Taylor instabilities \citep{stri2000b}.
The resulting escape of both pristine and enriched gas from the galaxy would
be especially dramatic for low-mass galaxies, due to their smaller escape
velocities.
However, if a hot and dense ICM is present, the wind is expected to grow more
slowly, perhaps even stall, and the swept-up shell not to immediately 
fragment due to its higher density \citep{mura1999}. 
This might leave a slowly cooling bubble of hot gas of
significantly higher density $n$ compared to a freely escaping wind
(resulting in a higher wind X-ray luminosity, since $L_X \propto n^2$).
In order to explain the metal abundance and extended star formation observed
in many low-mass galaxies,
some studies (e.g.\ Legrand et al.\ 2001) invoke the presence of
extended gaseous galactic haloes, acting as a barrier to the escape of
starburst winds.
The validity of this explanation could be questioned for low-mass galaxies in
groups, where such haloes, after all, might be stripped. 

Models and hydrodynamical simulations have been used to investigate 
the effect of an ambient ICM on the properties of a galaxy wind
(\citealt{mura1999}; \citealt{sili2001}), indicating that for sufficiently 
high ICM thermal pressures 
the ICM can indeed provide a confining mechanism. However, little effort has
been invested in quantifying the conditions required for ICM confinement. 
Furthermore, since no X-ray observations have been reported of
starburst galaxies embedded in the hot ICM of groups
or clusters, apart from the very low density gas in the compact group HCG16
\citep{turn2001}, such simulation results remain to be tested by 
direct observations.

Starburst winds may also be responsible for ``preheating'' the ICM in 
groups and clusters, most notably evidenced by the entropy profiles of hot gas
in groups (e.g.\ \citealt{ponm1999}; \citealt{sun2003}; \citealt*{ponm2003};
\citealt {rasm2004a}),
and they may even be capable of sweeping out a non-negligible fraction of the 
ICM from the most loosely bound systems \citep*{davi1999}. 
Further, some combination
of winds and stripping is expected to be responsible for the observed
chemical enrichment of the ICM. There is some
controversy, however, as to the ICM metallicity of groups, and very low
values compared to those found in clusters have been reported
(e.g.\ \citealt{davi1999}).
If true, this could imply that galactic winds in
some cases {\em are} suppressed by environment, a possibility of potential
relevance also to clusters, since groups act as cluster precursors in
hierarchical structure formation scenarios. If gas loss 
from galaxies is suppressed in groups, the metals in the cluster ICM may 
have been put in place more recently than is often supposed. 

In an attempt to shed light on these issues and clarify the role of the 
ICM in the late stages of a starburst wind, we obtained {\it Chandra} data of
the dwarf starburst galaxy NGC\,1800, located in galaxy group no.\ 62 in the 
catalog of \citet*{maia1989}. 
NGC\,1800 was selected on the basis of its proximity, its relatively small
distance from the centre of the group in which it is embedded 
(referred to as MdCL62 in the following), and because 
it is well inclined to our line of sight, thus allowing an efficient 
identification of sources in the disc and a clear view of any hot gas 
expanding out of the galactic plane.
The galaxy was detected in the {\em IRAS} Survey, 
showing 60 $\mu$m and 100 $\mu$m fluxes 
of $S_{60}=0.79$ Jy and $S_{100}=1.75$ Jy, hence displaying a warm 
far-infrared (FIR) 'temperature' ($S_{60}/S_{100}=0.45$), 
indicative of starburst activity. It is, indeed, recognized as a 
{\em bona fide} starburst (e.g.\ \citealt{calz1995}). It  
further seemed a particularly attractive target for a study of this type,
given that H$\alpha$ observations have revealed filamentary emission
extending well above the disc midplane, consistent with the picture that 
hot gas is being funneled above the disc \citep{hunt1996}. In this case there 
may be evidence that the swept-up shells have not fragmented completely and 
that the H$\alpha$ emission marks the edge of a stalled or slowly moving wind.
Additional motivation for studying NGC\,1800 was provided by the fact that 
only a handful of active dwarf starbursts 
have been studied in detail in X-rays, so the generic X-ray properties of this
class of objects remain poorly established. 
Furthermore, since dwarf galaxies are among the earliest objects to form (and
produce stars) in hierarchical structure formation scenarios, the study of 
nearby dwarfs may provide clues to the importance of different 
processes acting during the earliest stages of structure formation in 
the Universe.

As is the case for many dwarf galaxies, the exact Hubble type of NGC\,1800 is 
not well-determined.
It is most likely a spiral (although it has been claimed to
be irregular, see e.g.\ Hunter 1996), possibly containing a bar, and possibly 
of Magellanic type. Here we take it to be of SBm type, consistent with
its stellar mass-to-light ratio of $M/L_B \approx 3$ (\citealt*{gall1981};
typical of larger spirals but near the upper extreme of irregulars).
The optical extent can be conveniently parameterized through $D_{25}$, the 
ellipse outlining a $B$-band isophotal level of 25 mag arcsec$^{-2}$. 
The galaxy group MdCL62, of which NGC\,1800 is a member, contains five other 
galaxies, none of which are covered by this {\it Chandra} pointing. 
Some characteristics for the galaxy and the group are listed in 
Table~\ref{tab,tab1}.

Section~\ref{sec,obs} deals with details of the observation and analysis, 
Section~\ref{sec,res} presents results for the X-ray emission of NGC\,1800 
and the embedding group, and in Section~\ref{sec,dis} we discuss the 
properties and possible fate of the NGC\,1800 starburst wind. 
Section~\ref{sec,comparison}
compares NGC\,1800 to other dwarf starbursts, and 
discusses the scaling properties of starburst galaxies 
down into the dwarf regime. Summary and conclusions are presented in 
Section~\ref{sec,summary}.
$H_0=75$ km s$^{-1}$ Mpc$^{-1}$ is assumed throughout. The distance of 
NGC\,1800 is then 7.4 Mpc \citep{tull1988}, and 1 arcmin corresponds to 
$\sim 2.2$ kpc.

\begin{table*}
\centering
 \begin{minipage}{127mm}
\caption{\protect{General properties of the NGC\,1800 galaxy
and the MdCL62 group.
Absolute $B$ magnitude and $D_{25}$ (major and minor diameter) 
characteristics are mean data from the Lyon-Meudon Extragalactic Database 
(LEDA), 
$\theta_{D25}$ is the $D_{25}$ major axis position angle north eastwards;
inclination (the angle of the polar axis with respect to the line
of sight) taken from the NASA/IPAC Extragalactic Database (NED);
group velocity dispersion $\sigma$ taken from the catalog of 
Maia et al.\ (1989) and dynamical mass $M$ from \citet{gall1981}; 
projected distance $R_{proj}$ between NGC\,1800 and the 
group centre computed from the angular offset.}}
\label{tab,tab1}
\begin{tabular} {cccccccc} \hline 
\multicolumn{1}{c}{Name} &
\multicolumn{1}{c}{RA} &
\multicolumn{1}{c}{Dec} &
\multicolumn{1}{c}{Hubble type} &
\multicolumn{1}{c}{$M_B$} &
\multicolumn{1}{c}{$D_{25}$} &
\multicolumn{1}{c}{$\theta_{D25}$} &
\multicolumn{1}{c}{Incl.} \\
 & (J2000) & (J2000) & & & (arcmin) & (deg) & (deg) \\ \hline
NGC\,1800 & 05 06 25.4 & $-31$ 57 15 & SBm & $-16.86$ & (2.00, 1.15) & 106  & 64 \\
\hline \\
\end{tabular}

\centerline{
\begin{tabular} {ccccccc} \hline 
\multicolumn{1}{c}{Name} &
\multicolumn{1}{c}{RA} &
\multicolumn{1}{c}{Dec} &
\multicolumn{1}{c}{$N_{gal}$} &
\multicolumn{1}{c}{$\sigma$} &
\multicolumn{1}{c}{$M$} &
\multicolumn{1}{c}{$R_{proj}$} \\
 & (J2000) & (J2000) & &(km s$^{-1}$)  & (M$_\odot$)  & (kpc) \\ \hline
 MdCL62 & 05 05 41.0 & $-31$ 53 00 & 6 & 260  & $6\times 10^{13}$ & 20\\
\hline
\end{tabular}
}
\end{minipage}
\end{table*}

\section{Observation and analysis}\label{sec,obs}

NGC\,1800 was observed by {\it Chandra} (obs.\ ID 4062) with the ACIS-S3 chip 
as aimpoint, for an effective exposure time of 46.2 ks, and with the CCD's at
a temperature of $-120^\circ$~C. The presumed group 
centre was situated on the I2 chip. 
Data were telemetered in Very Faint mode which 
allows for superior background subtraction relative to standard Faint mode. 
To exploit this, the data were reprocessed and background screened using 
the {\tt acis\_process\_events} tool in {\sc ciao} v2.21.
Bad pixels were screened out using the bad pixel map provided by the pipeline,
and remaining events were grade filtered, excluding ASCA grades 1, 5, and 7.

For the analysis of NGC\,1800 we considered events on the S3 chip only.
Periods of high background 
were filtered using $3\sigma$ clipping of the 2.5--7 keV 
lightcurve extracted from this chip in 200 s bins, excluding bright point 
sources and a 2 arcmin diameter circle centred on the (optical) centre of 
the galaxy. The cleaning level resulting from this approach is potentially 
sensitive to the adopted bin size, since 
$\sigma \simeq (\mbox{counts bin$^{-1}$})^{1/2}$; we found, however, that the 
cleaned exposure time stayed constant to within 2 per cent for bin sizes 
100--1000 s. Some flares in the resulting lightcurve are 
evident, and removing the affected periods leaves a total of 34.6 ks of 
useful exposure time.

Point sources were identified on S3 using the wavelet-based {\sc ciao} tool 
{\tt wavdetect},
adopting a threshold significance of $1.8\times 10^{-6}$, which should limit
the number of spurious detections on the chip to $\sim 1$.
A total of 36 sources were detected, in fair agreement with
the statistical expectation of $\sim 25$ background sources for a 35 ks 
pointing (cf.\ \citealt{summ2003}).
We also tried {\tt vtpdetect} and {\tt celldetect} with default settings, but 
both methods seem to miss the fainter point sources, so the {\tt wavdetect} 
results were taken to be the most reliable. 
The detection run was repeated with lower thresholds to test if sources just 
below the adopted threshold were present; this was not found to be the case. 
A total of three point sources were detected inside $D_{25}$, with two more 
just outside, some of which may of course be unassociated with NGC\,1800. 
Unless otherwise specified, all sources were masked out in all further 
analysis, using their $3\sigma$ detection ellipses from {\tt wavdetect}.

Fig.~\ref{fig,image1} shows a raw soft-band image of the full S3 chip. 
It is immediately clear that very few source counts are present within 
the $D_{25}$ ellipse.
The other ellipse shown in the figure marks the region adopted for spectral 
extraction, hereafter denoted $D_{spec}$, centred at 
$(\alpha,\delta)_{2000} = (05^h06^m25\fs 5, -31\deg 57\arcmin22.5\arcsec)$. 
Diffuse emission seems confined to this region, as will be demonstrated 
below. Rectangles labelled 'A', 'B', and 'C' in Fig.~\ref{fig,image1} 
represent the regions adopted for spectral background estimation. 
Using regions in the data themselves for this purpose rather than publicly 
available blank-sky fields seems preferable,
given the small number of source counts, the limited extent of diffuse 
emission, 
and the potential contamination from group emission on the chip. 
Three partially overlapping regions, sized 5.5--9.3 arcmin$^2$, were selected
for background estimation, ensuring they were chosen neither too small, nor
too large, in order to limit sensibility to statistical errors and to
suppress systematics resulting from any significant response variations 
across the chip. We found no statistically 
significant differences between the area-weighted mean count rates of the 
three background regions or in resulting best-fitting model parameters for 
background subtracted source spectra.
Spectra were accumulated in bins containing at least 15 
net counts and fitted assuming $\chi^2$ statistics. 
All spectral results are presented for background region 'A'. 

\begin{figure*}
\begin{center}
\mbox{\hspace{-0.3cm}
\epsfxsize=13cm
\epsfysize=10.5cm
\epsfbox{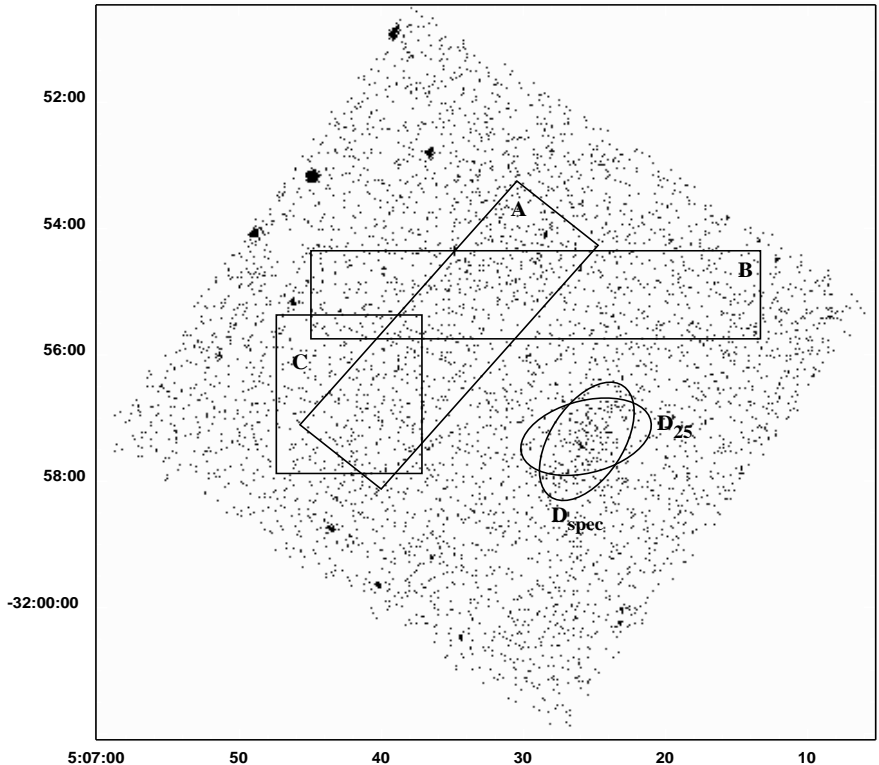}\hspace{0cm}}
\caption{Raw 0.3--2 keV image 
(spatial scale 1.5 arcsec pixel$^{-1}$) showing the regions used for 
extraction of source spectra ($D_{spec}$ ellipse), background spectra 
(rectangles),
and the $D_{25}$ ellipse determined from $B$-band isophotes. 
North is up and east is to the left.}
\label{fig,image1}
\end{center}
\end{figure*}

\section{Results}\label{sec,res}

\subsection{NGC\,1800 diffuse emission}\label{sec,diffuse}

Images were produced in various energy bands with 1 arcsec pixels in order to 
search for diffuse emission in and around NGC\,1800.  
To increase the signal-to-noise ratio (S/N) and highlight any such emission 
we employed a 
background reduction approach analogous to that of \citet{fabi2003}:
Photons were removed if they had less 
than three neighbours inside a radius of 3 arcsec, resulting in a 
uniform and substantial reduction outside source regions across the chip. This
method only served to enhance the contrast of source emission against the
background, and the resulting images were used solely for display purposes 
and not for quantitative analyses such as the extraction of surface brightness
profiles described below.
Fig.~\ref{fig,image2} shows a background reduced 0.3--5 keV 
image of the central $\sim 3\times 3$ arcmin$^2$ region around $D_{25}$,
adaptively smoothed using the {\sc ciao} task {\tt csmooth}. 
Diffuse emission is clearly seen, along with four of the five point 
sources detected by {\tt wavdetect} within or close to $D_{25}$ 
(the faintest point source, labelled '2' close to the S edge of $D_{25}$ 
and containing only four 0.3--5 keV counts, is not picked up by {\tt csmooth}
in this case).
The diffuse emission seems elongated in the northwest--southeast direction, 
confined inside an ellipse which is tilted at an angle
of $\sim 45 \deg$ relative to the major axis defined by $D_{25}$. 
The emission extent is 
roughly $1.1 \times 2.1$ arcmin$^2$ (ellipse major axes), corresponding to an 
enclosed area inside $D_{spec}$ of $\sim 9$ kpc$^2$. To the north of the 
galaxy centre, the tip of the X-ray emission coincides with the central region
of a web-like structure of H$\alpha$--emission detected by Hunter (1996), 
the approximate position and extent of which is outlined by a dashed 
rectangle in the figure. To the south, a H$\alpha$ shell was detected
(marked as a dashed line), approximately where $D_{25}$ intercepts 
the X-ray emission. 

\begin{figure*}
\begin{center}
\mbox{\hspace{-0.8cm}
\epsfxsize=10cm
\epsfysize=7.5cm
\epsfbox{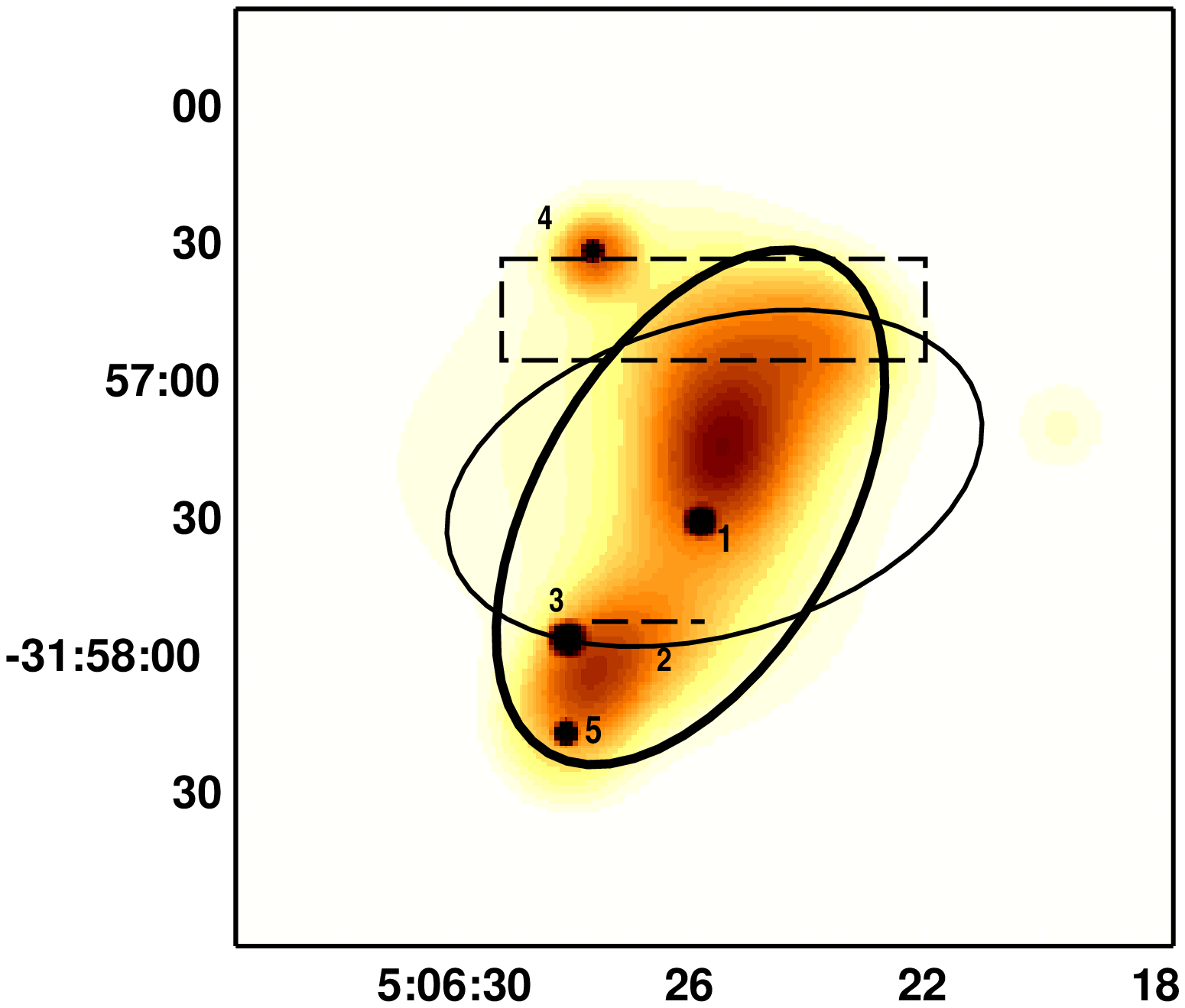}\hspace{0.0cm}
\epsfxsize=7cm
\epsfysize=7cm
\epsfbox{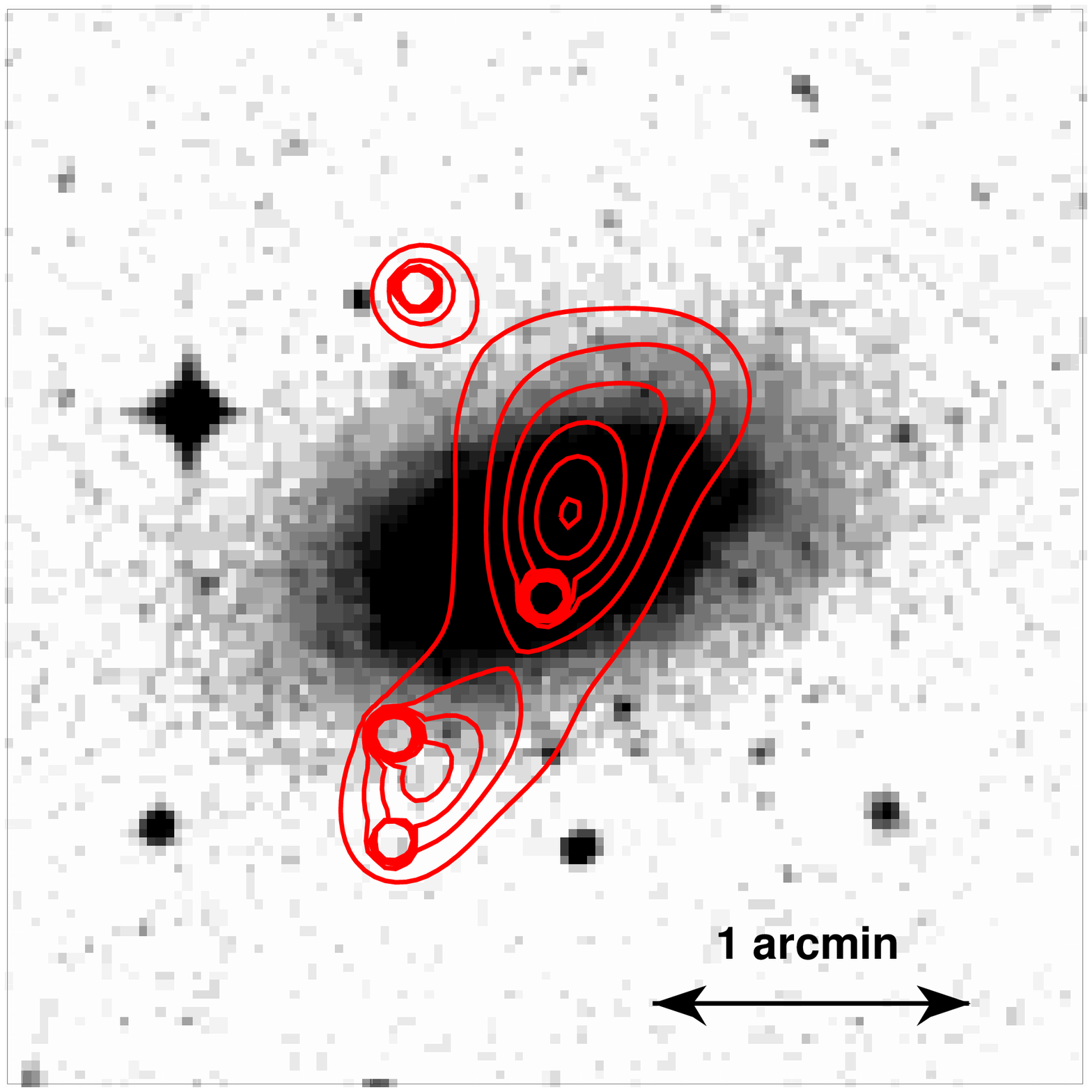}}
\caption{{\em Left}: Background reduced and adaptively smoothed 
0.3--5 keV image ($3\sigma-5\sigma$ significance range) displaying the 
central region around NGC\,1800. Ellipses mark the same
regions as in Fig.~\ref{fig,image1}. Dashed box and line mark the H$\alpha$
features detected by Hunter (1996), see text. Numbers indicate the five point
sources listed in Table~\ref{tab,pointsources} below. {\em Right}: 
Adaptively smoothed X-ray contours of the left figure overlayed on a 
Digitized Sky Survey optical image of NGC\,1800.}
\label{fig,image2}
\end{center}
\end{figure*}

Inspection of images produced with larger binning revealed no evidence for 
low-surface brightness emission surrounding the source region.
To investigate the surface brightness structure of the diffuse emission, 
we extracted the counts in a rectangular aperture with the same centre, 
position angle, and width as the $D_{spec}$ ellipse of 
Figs.~\ref{fig,image1} and \ref{fig,image2}. 
The rectangle height was chosen such as to extend well beyond
the $D_{25}$ ellipse, in order to further test for the presence of diffuse
emission outside $D_{25}$. The row-averaged profile is shown in 
Fig.~\ref{fig,surfbright2}, binned in 8 arcsec bins. The low number of counts
precludes any firm conclusions regarding the shape of the profile, and thus
the density distribution and 3-D morphology of the wind, but
this figure clearly shows little evidence for diffuse 
emission outside $D_{spec}$ where the profile is lost in the background.
The lumps seen roughly 100 arcsec away at either side of the $D_{spec}$ 
centre are not highly significant, and could not be associated with obvious
features in the raw or smoothed images, or with detector artefacts when 
viewing the data in detector coordinates. 
\begin{figure}
\begin{center}
\mbox{\hspace{-0.6cm}
\epsfxsize=9.3cm
\epsfysize=7.3cm
\epsfbox{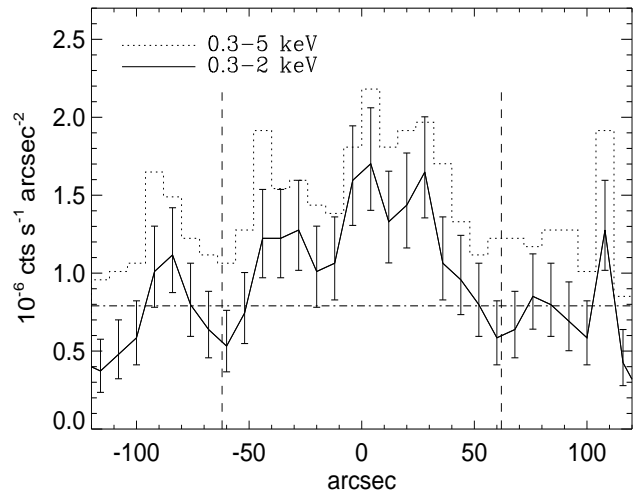}\hspace{0cm}}
\caption{Surface brightness profiles of diffuse emission in a
rectangular region centred on the $D_{spec}$ ellipse, the extent of which
(major axis) is shown by vertical dashed lines. $1\sigma$ error bars shown 
for the 0.3--2 keV profile have been calculated using the expressions of 
\citet{gehr1986} for small-number Poissonian errors. 
Horizontal dash-dotted line marks the 0.3-2 keV background 
level as estimated from region 'A' in Fig.~\ref{fig,image1}. The northern
side of the galaxy is to the right.}
\label{fig,surfbright2}
\end{center}
\end{figure}

For the extraction of spectra of the diffuse emission we experimented with 
different apertures,
finally adopting the $D_{spec}$ region shown in Figs.~\ref{fig,image1} and 
\ref{fig,image2} on the basis that it maximizes the S/N ratio while
ensuring that nearly all source counts are included
(increasing the size of the region does not increase the number of net
counts, and extracting spectra e.g.\ inside $D_{25}$ instead, yields $\sim 
30$ per cent fewer net counts over roughly the same detector area).
Spectra of the diffuse emission were first extracted inside $D_{spec}$ using
responses weighted by the observed number of counts in the 0.3--3 keV band.
The {\sc acisabs} model\footnote{Available from the {\sc ciao} web pages
published by the {\em Chandra} X-ray Center, 
http://cxc.harvard.edu/ciao/} 
was applied to the effective area file prior to 
spectral fitting, to account for the low-energy degradation in 
quantum efficiency of the ACIS chips. 
Only $\sim 110$ net counts from diffuse
emission were detected
inside $D_{spec}$. A good fit to the spectrum could nevertheless be obtained 
using an absorbed 
{\sc mekal} model (reduced $\chi^2=0.84$) in {\sc xspec} v11.3, whereas an 
absorbed power-law provided a poorer description 
(reduced $\chi^2 = 1.7$). 

Preferably, responses should be weighted using the {\it incident} photon 
spectral distribution rather than the observed one 
(since the net effective area varies significantly with energy in the 
0.3--3 keV range). 
Our best guess for the incident spectrum is the above 
best-fitting model, so spectra were re-extracted with this model acting as 
spectral weights. This, however, produced negligible differences with respect 
to the first result.
The resulting spectrum is shown in Fig.~\ref{fig,diffusespec}.
An absorbed {\sc mekal} model was re-fitted to the spectrum. Given the small 
number of source counts and the ACIS 
low-energy calibration uncertainties, we chose to fix the absorbing column
density $N_H$ at the Galactic value of $1.54\times 10^{20}$ cm$^{-2}$
(with which it was indeed found to be consistent), while
$Z$ was initially fixed at $0.5$Z$_{\odot}$, assuming solar abundances from 
\citet{ande1989}. An absorbed power-law was also 
fitted, and Table~\ref{tab,spec} lists the results of spectral fitting.
A differential emission measure model ({\sc cemekl} in {\sc xspec}) 
failed by a large margin to reproduce the flat spectral shape between 
0.5 and 1 keV and is not included in the Table.

\begin{figure}
\begin{center}
\resizebox{8.4 cm}{!}{\rotatebox{-90}{\includegraphics{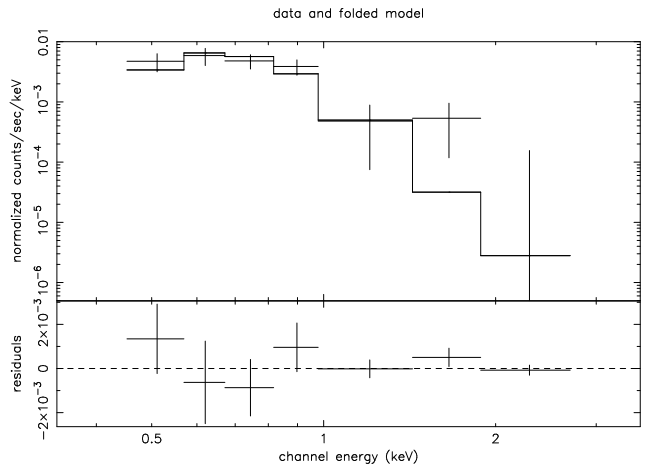}}}
\caption{\protect{0.3--3 keV spectrum of NGC\,1800 diffuse emission,
and best-fitting {\sc mekal} plasma model. Bottom plot shows fit residuals.}}
\label{fig,diffusespec}
\end{center}
\end{figure}

\begin{table*}
\centering
 \begin{minipage}{104mm}
  \caption{\protect{Summary of spectral fits to the 0.3--3 keV 
diffuse emission. All errors are $1\sigma$.}}
   \label{tab,spec}
\begin{tabular} {lll} \hline
\multicolumn{1}{l}{Model} &
\multicolumn{1}{c}{Fit results} &
\multicolumn{1}{c}{Goodness of fit}\\  \hline \vspace{0.15cm}
{\sc abs(mekal)}   & $T = 0.24^{+0.03}_{-0.02}$, $Z = 0.5Z_{\odot}$ (fixed) & $\chi^2=4.2/5$ (= 0.84)  \\ \vspace{0.15cm}
{\sc abs(mekal)}   & $T = 0.26^{+0.05}_{-0.03}$, $Z = 0.05^{+0.22}_{-0.04} Z_{\odot}$ & $\chi^2 = 3.1/4$ (= 0.77) \\ 
{\sc abs(powerlaw)}& $\Gamma = 3.78^{+0.44}_{-0.40}$ & $\chi^2=8.7/5$ (= 1.73)\\
\hline \\
\end{tabular}
\end{minipage}
\end{table*}

As can be seen from Table~\ref{tab,spec}, a thermal {\sc mekal} model 
provides a good fit, whereas a power-law is not a good description of the 
data. The low number of counts does not warrant the introduction of additional
model components for spectral fitting, so whether a second thermal
component with $T\sim 0.6-0.8$ keV is present as in 
certain other dwarf starbursts (e.g.\ \citealt{summ2003}; \citealt{summ2004};
\citealt{hart2004}) cannot be addressed with the present data. 

The metal abundance, when left as a free parameter in the fits,
is not well constrained. It is consistent with the (low) values of 
$Z_{\mbox{\small Fe}}\approx 0.05$Z$_{\odot}$ found for the Fe abundance of 
the soft thermal component in the starburst winds of NGC\,253 and M82
\citep{ptak1997}, and with results from optical spectroscopy of the three
metal-deficient ($Z\approx 0.02$Z$_{\odot}$) star-forming blue compact dwarfs 
studied by \citet{thua2004}. It also 
marginally agrees with estimates of the oxygen abundance in the ionized ISM
of NGC\,1800, $0.2$Z$_{\odot}<Z_{\mbox{\small O}}<0.4$Z$_{\odot}$ 
\citep{gall1981}, and with X-ray results for the winds in the dwarf starbursts
NGC\,1569 ($>0.25$Z$_{\odot}$; \citealt*{mart2002}), NGC\,4449 ($0.32\pm 0.08$Z$_{\odot}$; \citealt{summ2003}), and NGC\,5253 ($0.14\pm 0.01$Z$_{\odot}$;
\citealt{summ2004}).
While the metallicity of the ISM in dwarf galaxies as inferred from optical 
spectroscopy is usually agreed to be fairly low ($\la$ 0.2--0.3 Z$_\odot$), 
there is a controversy as 
to the amount of chemical enrichment of any outflowing gas in these galaxies
(see e.g.\ \citealt{tosi2004}). The picture suggested by the above 
observations is that dwarf starburst winds are not substantially 
enriched but may have metallicities comparable to the ISM. This is in line
with some analytical models for the chemical evolution of dwarf galaxies
(e.g.\ \citealt*{lars2001}) and with the idea that the expanding shells
sweep up a significant portion of the surrounding ISM material. In fact, 
hydrodynamical simulations \citep{stri2000b} suggest
that wind abundance determinations from X-ray spectroscopy may reflect 
the values in disc and halo ISM to a substantial degree, but also that X-ray 
observations do not directly probe the global metallicity of winds.
Furthermore, such simulations indicate the presence of a complex multi-phase
environment with gas at a range of temperatures, and there is a well-known 
bias towards low derived abundances when a multi-temperature gas is fitted 
with a single-temperature model (\citealt{buot2000}; \citealt{buot2003}).
Consequently, the above results should be interpreted with some caution.

From the best-fitting model, with $Z$ fixed at its best-fitting value of 
$0.05$Z$_{\odot}$, the resulting unabsorbed 0.3--3 keV flux from diffuse
emission is 
$1.9\pm 0.5 \times 10^{-14}$ erg cm$^{-2}$ s$^{-1}$, with errors 
from the fractional $1\sigma$ errors on the spectral normalization.
At the distance of NGC\,1800, this translates into a total diffuse luminosity 
of $1.3\pm 0.3 \times 10^{38}$ erg s$^{-1}$.

The number of source counts from the five point sources detected within or 
close to $D_{25}$ totals less than 90 (0.3--5 keV), so any detailed 
investigation of point source X-ray properties is excluded. 
A power-law fit to their combined spectrum 
suggests a photon index of $\Gamma \approx 0.8$ and a combined 0.3--3 keV flux
of $1.2\pm 0.2 \times 10^{-14}$ erg cm$^{-2}$ s$^{-1}$. Assuming the sources 
to be at the distance of NGC\,1800, this would imply a combined 0.3--3 keV
luminosity of $\approx 8\times 10^{37}$ erg s$^{-1}$. 
Resolved point sources thus account 
for $\sim 40$ per cent of the observed soft X-ray emission, a value 
bracketed by the corresponding results for other dwarf starbursts (e.g.\ 
\citealt*{ott2003}; \citealt{summ2003}).
These results are dominated by the two point sources outside $D_{25}$, 
however. The hardness of the best-fitting spectrum may
suggest that these sources are intrinsically absorbed active galactic nuclei 
(AGN) and therefore 
are unassociated with NGC\,1800. Fitting these two sources alone produces a 
$\Gamma$ of $\sim 0.6$. Table~\ref{tab,pointsources} lists some 
properties of the five point sources. As can be seen from 
Fig.~\ref{fig,image2}, none of these X-ray point sources have a bright 
optical counterpart, and it is not possible to associate them with 
near-infrared point sources or quasars detected in the 
2 Micron All Sky Survey (2MASS). 
The inferred luminosities are representative of X-ray
binaries, and we find no evidence for the presence of ultraluminous X-ray
sources, with $L_X \sim 10^{39} - 10^{41}$ erg s$^{-1}$.

\begin{table*}
\centering
\begin{minipage}{106mm}
\caption{\protect{Properties of the five point sources detected
within or very close to the $D_{25}$ ellipse, numbered according to increasing
projected distance from the $D_{25}$ centre (cf.\ Fig.~\ref{fig,image2}). 
Background-subtracted count rates and luminosities are given
in the 0.3--5 keV band, the latter values assuming a power-law spectrum of 
$\Gamma=0.8$ from a fit to the combined spectrum. All errors are $1\sigma$.}}
\label{tab,pointsources}
\begin{tabular} {cccrc} \hline 
\multicolumn{1}{c}{Source no.} &
\multicolumn{1}{c}{RA} &
\multicolumn{1}{c}{Dec} &
\multicolumn{1}{c}{Count rate} &
\multicolumn{1}{c}{$L_X$} \\  
  & (J2000) & (J2000)                   & (counts s$^{-1}$) $ $ $ $ $ $& ($10^{37}$ erg s$^{-1}$) \\ \hline 
1 & 05 06 25.85 & $-31$ 57 27.7 &  $3.3\pm 1.0\times 10^{-4}$ & $1.9\pm 0.6$\\
2 & 05 06 26.95 & $-31$ 57 52.7 &  $1.3\pm 0.6\times 10^{-4}$ & $0.7\pm 0.3$\\
3 & 05 06 28.12 & $-31$ 57 52.8 &  $3.6\pm 1.0\times 10^{-4}$ & $2.0\pm 0.6$\\
4 & 05 06 27.74 & $-31$ 56 28.7 & $14.2\pm 0.2\times 10^{-4}$ & $7.9\pm 1.2$\\
5 & 05 06 28.17 & $-31$ 58 13.3 &  $7.2\pm 1.5\times 10^{-4}$ & $4.0\pm 0.8$\\ \hline 
\end{tabular}
\end{minipage}
\end{table*}

\subsection{A hot gaseous halo around NGC\,1800?}

There is clear evidence from H{\sc i} observations for gas extending well 
beyond the optical extent of the galaxy (out to twice the Holmberg radius
$D_{26.5}$; \citealt*{hunt1994}), and the presence of a diffuse H$\alpha$ 
halo has been claimed by \citet{hunt1996}.  
Numerical simulations of disc galaxy formation (e.g. \citealt{toft2002})
further predict the presence of
a hot, X-ray emitting gaseous halo surrounding such galaxies.
NGC\,1800 is not a normal spiral, and, being possibly irregular, it might not 
even qualify as a disc galaxy. But a hot X-ray halo, fueled by
outflowing bubbles of gas, could also be expected on the basis of the 
starburst activity in NGC\,1800.
To test for the presence of an X-ray emitting halo around the galaxy,
a spectrum was extracted in an elliptic, 1.5 arcmin wide annulus outside 
the combined region of $D_{25}$ and $D_{spec}$.
We found no evidence for excess emission above the background in this
annulus. The predictions of \citet{toft2002}, 
when extrapolated to a disc circular velocity of
$V_c \approx 30$ km s$^{-1}$ as is relevant for NGC\,1800 
(Section~\ref{sec,fate}), would 
suggest $T \la 0.1$ keV and a 0.3--3 keV 
luminosity of $\sim 10^{37}$~erg~s$^{-1}$ for the halo gas. 
For a $T=0.1$ keV $Z=0.3$Z$_{\odot}$ {\sc mekal} plasma, the absence of a 
detection
inside the adopted annulus translates into a $3\sigma$ upper limit on the 
0.3--3 keV flux and luminosity of
$6.6\times 10^{-15}$~erg~cm$^{-2}$~s$^{-1}$ and
$4.2\times 10^{37}$ erg s$^{-1}$, 
respectively, in consistency with the above expectations.

A trivial reason for the lack of a hot halo detection is that such a
halo is simply too
X-ray faint to be detectable in these observations. This is surely a
realistic possibility, given the predictions of hot halo X-ray properties 
reported by \citet{toft2002} and the observational results of 
\citet{bens2000}. Alternatively, the standard picture of disc galaxies 
surrounded by a quasi-hydrostatic 
halo containing gas at the virial temperature is invalid for low-mass
systems and/or systems formed at high redshifts 
(see e.g.\ \citealt{birn2003}). Dwarf galaxies satisfy both these 
criteria. The possibility also exists that any hot 
halo has been ram-pressure stripped through interactions with the group ICM 
(cf.\ \citealt*{marc2003}). 
The presence of an extended H{\sc i} halo, as demonstrated by 
\citet{hunt1994}, suggests, however, 
that ram pressure effects are not significant.

\subsection{Group diffuse emission}\label{sec,group}

The group centre in the \citet{maia1989} catalogue 
(Table~\ref{tab,tab1}) is located on 
the I2 chip, but given that the group is rather sparse, the centre of the
group potential may not be well-determined from optical observations.
To give an appreciation of the nature of the group and the angular scales 
involved, we show in Fig.~\ref{fig,n1800group} the sky position of the
optically determined group centre and the six member galaxies.
\begin{figure*}
\begin{center}
\mbox{\hspace{-0.6cm}
\resizebox{11cm}{!}{\includegraphics{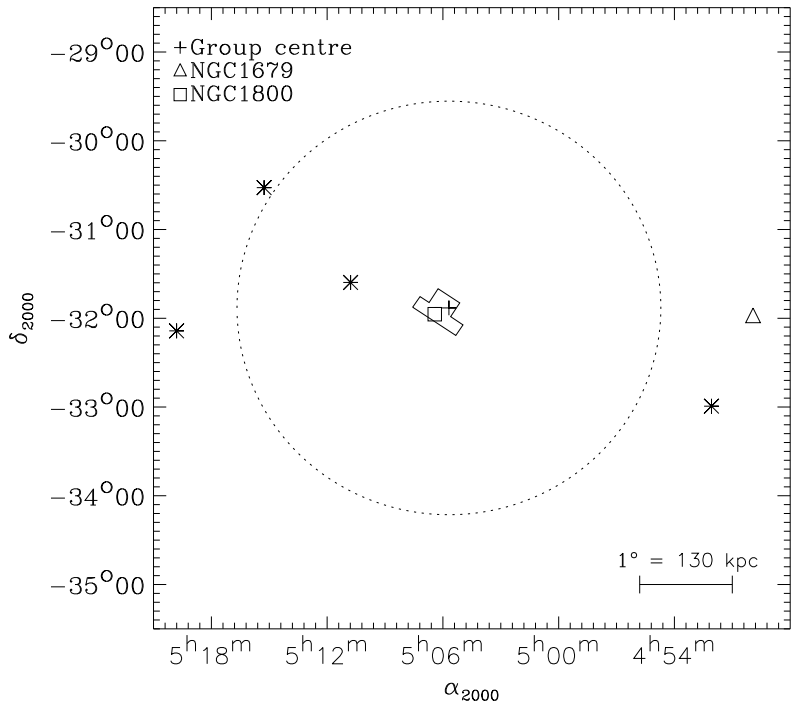}}}
\caption{\protect{Galaxy distribution in the MdCL62 group
in a $7\times 7$ deg$^2$ region.
A cross marks the optically determined group centre, NGC\,1800 is shown by a 
square, and the brightest group galaxy, NGC\,1679, by a triangle. Remaining
galaxies are represented by asterisks. Also 
outlined is the area covered by the ACIS CCD's (solid) for this observation,
and the radius $r=300$ kpc (dotted) used for estimating the X-ray 
luminosity of intragroup gas (see text).}}
\label{fig,n1800group}
\end{center}
\end{figure*}
There is no clear indication of the position of the group centre in 
the present X-ray data, so in order to search for X-ray emission from 
intragroup
gas, we decided to extract large-area spectra of source-free regions on both
the I2 and S3 chips. For background estimation, we employed
the 'period~D' blank-sky data of 
M.\ Markevitch\footnote{http://hea-www.harvard.edu/$\sim$maxim/axaf/acisbg/}, 
screened and filtered
as for our data and taking advantage of the fact that the blank-sky 
data were also made in Very Faint mode. The $\sigma$--$T$ relation of 
\citet{hels2000} would suggest $T\approx 0.7$ keV for the intragroup 
gas, so any group X-ray emission is likely to be fairly soft. Differences in 
local soft X-ray background between source and background data 
could thus potentially influence the results. We therefore checked 
the exposure-weighted mean value of {\it ROSAT} All-Sky Survey soft 
(R45 band, $\sim 0.5$--0.9 keV) count rates in the two data sets, finding
$1.01\pm 0.10 \times 10^{-4}$ 
(for a $2\times 2$ deg$^2$ region around NGC\,1800) and 
$1.13\pm 0.11 \times 10^{-4}$ cts s$^{-1}$ arcmin$^{-2}$ 
for source and background data, respectively, and hence no significant 
difference.

For the I2 chip, we performed a $3\sigma$ clipping of the data based on the 
0.3--12 keV 
lightcurve, and excluded, from source and background data, the source regions
found by {\tt wavdetect} in source data. 
The blank-sky data were subsequently coordinate transformed to match the
pointing characteristics of our source data, ensuring that source and
background spectra could be extracted from identical detector regions.
It was then confirmed for the S3 chip that the
blank-sky background did not differ significantly from our local background 
estimates, by extracting a source and background spectrum inside $D_{spec}$ 
and performing a fitting procedure analogous to that described in
Section~\ref{sec,diffuse}. Results were in excellent agreement with those 
obtained using local background determinations 
(rectangles 'A'--'C' in Fig.~\ref{fig,image1}), 
including the resulting number of net counts inside $D_{spec}$. 
This in itself suggests that there is little group emission from gas along
our line of sight to NGC\,1800.

To search for group diffuse emission well outside $D_{25}$, a 'full-chip' 
spectrum was extracted, avoiding chip edges and excluding sources
from {\tt wavdetect} and the region covered by twice the ellipse axes of 
$D_{25}$ (equivalent to the Holmberg radius of the galaxy, cf.\ 
\citealt{hunt1994}). Within the $1\sigma$ background noise we found no
evidence for excess emission in the 0.3--5 keV band on
S3. The exercise was repeated for the I2 data, with the same result. 

The MdCL62 group was not detected in the {\em ROSAT} All-Sky Survey, 
which features an exposure time of $\approx 500$~s at this sky position. 
The $L_X$--$\sigma$ relation of \cite{hels2000}, based on {\em ROSAT} 
data of 24 X-ray bright groups, would suggest a (bolometric)
luminosity of $L_X \approx 1.5 \times 10^{42}$ erg s$^{-1}$.
Assuming the emission to be uniformly distributed within 300 kpc 
(typical of the group radii probed by {\it ROSAT}), centred on the optical 
group centre, 
with $T=0.7$ keV, $Z=0.2$Z$_{\odot}$, and subject to the relevant value of 
$N_H$, a {\sc pimms} calculation suggests that we should expect 
$\sim 1300$ and $\sim 1800$ counts on the I2 and S3 chips, 
respectively (0.3--5 keV, corrected for removed area). 
Although not corrected for vignetting, these estimates should be 
lower limits, as the true X-ray surface brightness distribution 
is expected to peak close to the observed region and thus should be well 
above the average value inside $\sim 300$ kpc on either chip. 
Even if the X-ray group centre is located 300 kpc {\it away} from the 
nominal aimpoint of this 
observation, with a surface brightness distribution following a standard
$\beta$-model of $\beta =0.5$ and $r_c = 100$ kpc 
(based on $r_{500}\approx 600$ kpc from its dynamical mass of 
$6\times 10^{13}$ M$_{\odot}$, cf.\ Table~\ref{tab,tab1},
and $r_c \sim 0.15r_{500}$  from \citealt{sand2003}),
the above estimates are still more than an order of 
magnitude above the observed values. 
It therefore seems safe to assume that
$L_X < 10^{41}$ erg s$^{-1}$ for the intragroup gas, implying that
the group is rather X-ray faint. This result is perhaps not surprising, given
that all six group members are spirals or dwarf spirals, with the brightest 
member being the Scd spiral NGC\,1679 at $M_B= -18.89$. In contrast, the 
dominant galaxy in X-ray bright groups is almost always of early-type (e.g.\
\citealt{mulc2003}).
The result also implies that the group falls well below the nominal 
$L_X$--$\sigma$ relation of \citet{hels2000}.
This relation is subject to a good 
deal of scatter, though, so it would seem premature to conclude, on these
grounds, that $\sigma$ for the group might be artificially boosted by the 
inclusion of galaxies unrelated to the group. The galaxy configuration
suggests the group is somewhat unrelaxed, so perhaps the group is still only
approaching a state of sufficiently high density to produce significant
X-ray emission.

\section{Discussion}\label{sec,dis}

\subsection{Hot gas and star formation properties}\label{sec,hot}

The X-ray luminosity and overall morphology of the hot X-ray gas in NGC\,1800
is similar to that inferred for other nearby dwarf starburst galaxies, in 
which a hot, outflowing galactic wind has been detected. It is therefore
natural to assume that the diffuse X-ray emission in NGC\,1800 is the
signature of such a wind.
Using the derived gas parameters for NGC\,1800 one can assess the physical 
properties of the X-ray emitting gas, once the size of the X-ray emitting 
volume has 
been estimated. Fig.~\ref{fig,image2} clearly suggests that this volume is 
nonspherical, so here it is assumed that the gas is confined inside an 
ellipsoid with two of the axes equal to those of $D_{spec}$ and the third 
(the one along our line of sight) equal to the $D_{spec}$ minor axis; 
this yields a total volume 
$V\approx 4.2 \times 10^{65}$ cm$^3$, of which only some fraction $\eta$ is 
occupied by X-ray gas. From the spectral normalization $A$ in {\sc xspec},
\begin{equation}\label{eq,xspec}
A = \frac{10^{-14}}{4\pi D^2} \int{n_e n_H \mbox{ d}V},
\end{equation}
where $D$ is the distance (Section~\ref{sec,intro}), one can derive the 
fitted 
emission measure $EM \equiv \int{n_e n_H \mbox{ d}V} \approx \eta n_e^2 V$, 
and hence the mean electron density $n_e \approx (EM/V\eta)^{1/2}$. 
This can then
be combined with $kT$ (Table~\ref{tab,spec}) and $L_X$  to infer the gas mass
$M_{gas} \approx m_p n_e V \eta^{1/2}$, thermal pressure 
$P\approx 2n_e kT \eta^{-1/2}$,
thermal energy $E_{th}\approx 3n_ekTV\eta^{1/2}$, cooling function 
$\Lambda = L_X/EM$, cooling time 
$t_{cool}\approx 3kT\eta^{1/2}/(\Lambda n_e)$, 
mass deposition rate
$\dot M_{cool}=M_{gas}/t_{cool}$, and the mean particle velocity 
$\langle v_p \rangle = (2E_{th}/M_{gas})^{1/2}$. Most of 
these quantities are listed in Table~\ref{tab,phys}. Unless otherwise stated,
we will in the following take $\eta =1$ for simplicity, and, where relevant,
discuss the implications of this assumption.

\begin{table*}
\centering
 \begin{minipage}{116mm}
\caption{\protect{Physical properties of the hot X-ray gas in
NGC\,1800. See text for the dependence of each quantity on the volume filling
factor $\eta$.}}
\label{tab,phys}
\begin{tabular}{ccccccc} \hline 
\multicolumn{1}{c}{$n_e$} &
\multicolumn{1}{c}{$M_{gas}$} &
\multicolumn{1}{c}{$P$} &  
\multicolumn{1}{c}{$E_{th}$} &  
\multicolumn{1}{c}{$t_{cool}$} & 
\multicolumn{1}{c}{$\dot M_{cool}$} & 
\multicolumn{1}{c}{$\langle v_p \rangle$} \\ 
(cm$^{-3}$) & (M$_{\odot}$) & (dyn cm$^{-2}$) & (erg) & (Myr) & (M$_{\odot}$ yr$^{-1}$) & (km s$^{-1}$) \\ \hline
0.012 & $4.1\times 10^6$ & $9.9\times 10^{-12}$ & $6.1\times 10^{54}$ & 1500 & 0.003 & 290\\
\hline \\
\end{tabular}
\end{minipage}
\end{table*}

The resulting mass of X-ray gas is well below the total neutral gas mass of 
NGC\,1800 of 
$\sim 2\times 10^8$ M$_{\odot}$, as inferred from H{\sc i} observations 
(\citealt{gall1981}; corrected to our assumed distance), and an order of 
magnitude lower than the mass of ionized gas in the H$\alpha$ web north of 
the galaxy ($\sim 4\times 10^7$ M$_{\odot}$; \citealt{hunt1996}). 
The mass and thermal energy content of X-ray gas is also smaller
than typical values inferred for non-dwarf starburst galaxies by
similar methods \citep{read2001}, but is
comparable to published values for other dwarf starbursts
(\citealt{mart2002}; \citealt{ott2003}; Summers et~al.\ 2003, 2004; 
\citealt{hart2004}).
The thermal pressure in the wind corresponds to $P/k \sim 7\times 10^4$
K cm$^{-3}$, an order of magnitude larger than the ISM pressure in the 
Galactic midplane \citep{wolf1995}. We should mention, though, that
results from hydrodynamical simulations \citep{stri2000b} indicate that
only a small fraction of the total energy and mass content of winds is 
probed by observations of their soft X-ray emission, with most of the thermal
and kinetic energy being carried by hot and rarefied low-emissivity gas 
which is difficult to detect observationally. Nevertheless, the above values
provide a first handle on the wind properties of NGC\,1800, and will enable 
a more detailed comparison to other starburst galaxies for which these 
quantities have been estimated using identical techniques 
(see Section~\ref{sec,comparison}).

The global star formation rate (SFR) of NGC\,1800 can be estimated from the
total H$\alpha$ luminosity of the galaxy.
For L(H$\alpha)\approx 4\times 10^{39}$ erg s$^{-1}$ (\citealt{hunt1996}; 
corrected to the assumed distance), the relation of \citet*{kenn1994}, 
SFR(M$_\odot$ yr$^{-1}$) = L(H$\alpha$)/($1.26\times 10^{41}$ erg s$^{-1}$), 
would suggest an SFR of $\approx 0.03$ M$_\odot$ yr$^{-1}$. 
For the adopted distance, this agrees well with earlier estimates by 
\citet{gall1981} and \citet{hunt1994}, also based on 
H$\alpha$ emission fluxes. The star formation rate can also 
be estimated from the (far-)infrared luminosity $L_{FIR}$ of the 
galaxy, assuming the relation of \citet{kenn1998},
SFR(M$_\odot$ yr$^{-1}$) = $L_{FIR}$/($2.2\times 10^{43}$ erg s$^{-1}$).
$L_{FIR}$ (8--1000 $\mu$m) is here estimated on the basis of IRAS 12, 25, 60,
and 100 $\mu$m fluxes using the expression of \citet{sand1996};
at 12 and 25 $\mu$m only upper limits are available, giving $L_{FIR}$
in the interval 1.2--2.6 $\times 10^8$ L$_\odot$, and hence 
SFR $\approx$ 0.02--0.04 M$_\odot$ yr$^{-1}$, in agreement with the
H$\alpha$ estimate.

The age of the current star formation activity can be constrained in a number
of ways. Based on the colours of the stellar population, \citet{gall1981} 
argue that this age should be less than $\sim 4\times 10^8$ yr, unless 
the pre-burst system was an intrinsically red inactive dwarf.
A lower limit 
to the age can be estimated if the H$\alpha$ filaments are interpreted as 
supershells that have cooled and 
fragmented. \citet{hunt1996} finds that the age of the bubbles should be at 
least 6 Myr and, more probably, tens of Myr, based on results from the  
hydrodynamical simulations of \citet*{macl1989}. A third estimate
can be obtained (cf.\ \citealt{heck2003}; \citealt{summ2004}) by assuming 
that the average mass injection rate into the 
X-ray bubbles during the starburst roughly corresponds to, or slightly 
exceeds, the {\em current} global star formation rate of 
$\approx 0.03$ M$_{\odot}$ yr$^{-1}$. 
The derived X-ray gas mass of $\approx 4\times 10^6$ M$_{\odot}$ 
then suggests a bubble age of $\sim 100$ Myr. 
Lower filling factors $\eta$ would imply a lower bubble age according
to this argument, but the filling factor for the NGC\,1800 wind cannot be 
reliably estimated from the present data. 
Simulations \citep{stri2000b} and {\it Chandra} observations 
of the NGC\,253 starburst \citep{stri2000a} suggest that the 
wind emission in these cases originates in low filling factor gas, for which 
$\eta < 50$ per cent, and possibly $\eta$ could be as low as a few per cent.
In the following we will consequently assume a bubble age of $\sim 50$ Myr.

\subsection{X-ray and H$\alpha$ morphology}

Although not a robust result due to the low number of detected counts,
Fig.~\ref{fig,image2} suggests the wind is an elongated bipolar outflow, 
possibly with a conical geometry. Any detailed discussion on the wind 
morphology is unfortunately not warranted by the data. 
It is, however, clear from Fig.~\ref{fig,image2} that the X-ray gas does 
extend beyond $D_{25}$, as do the observed H$\alpha$ filaments. Where the 
northern X-ray emission is lost in the background, a H$\alpha$ web is seen 
\citep{hunt1996}, running parallel to the $D_{25}$ major axis for 
$\sim 2$ kpc.
To the S, a H$\alpha$ loop is detected, interpreted by \citet{hunt1996} as a 
supershell extending away from the major axis, based on its size of 
$\sim 500$ pc and its shape, with the loop ends extending back towards the 
main body of the galaxy. In addition, the central regions feature a total of 
9--10 `fingers' of H$\alpha$ emission seen on either side of the galaxy, 
roughly aligned with the minor axis and extending away from the major axis 
for several hundred pc.
Apart from these features \citet{hunt1996} also detects a larger halo of 
diffuse ionized gas. 

\citet{hunt1996} suggested that both the northern H$\alpha$ web and the 
southern loop could be supershells that have expanded out of
the plane, cooled, and fragmented. The southern supershell may not yet have
cooled and fragmented completely, and could be younger than the N web but 
would eventually produce a similar structure. 
The emission-line ratios of the web were found to be consistent with 
direct photoionization by massive stars being the dominant ionization 
mechanism. The central H$\alpha$ fingers could be galactic chimneys,
providing the low optical depths required for UV photons from massive stars
in the disc to traverse the 1--2 kpc to the web. These chimneys could also 
be funneling hot, X-ray gas above the galactic plane.
We note that \citet{norm1989} find for a chimney 
model of the Milky Way that the chimney phase is associated with a mass flow 
rate of 0.3--3 M$_{\odot}$ yr$^{-1}$.
For NGC\,1800, these values are likely to be smaller and so probably 
consistent with the current star formation rate.
As noted by \citet{hunt1996}, the 
time-scales involved would then imply that the H$\alpha$ filaments and fingers
have been produced by a previous generation of massive stars rather than
the current generation responsible for the observed X-ray emission.

While the emission-line ratios of the H$\alpha$ web generally differ from 
those expected for
collisionally excited gas, it would seem natural to assess to which extent 
the H$\alpha$ gas may, in fact, be former hot X-gas from the 
outflowing bubbles that has now cooled to temperatures $\la 10^5$ K. 
With the estimated cooling time of X-ray gas ($t_{cool} \sim 1500$ Myr), 
the upper limit to the age of the current star formation activity of 
$\sim 400$ Myr 
suggests this is not a viable possibility. The even larger discrepancy 
between $t_{cool}$ and our estimate for the outflow age of a few tens of Myr 
further strengthens this conclusion. In fact, to produce the mass of the 
H$\alpha$ web ($\sim 4\times 10^7$ M$_\odot$) with the current mass 
deposition rate derived for the X-ray gas would require more than a Hubble 
time. It thus seems unlikely that both the X-ray and H$\alpha$ emission is 
due to radiative cooling of the wind fluid. This conclusion is similar to 
that inferred for the NGC\,253 starburst \citep{stri2002}, and agrees with 
the result of \citet{hunt1996} that the H$\alpha$ web is predominantly 
photoionized.
The flux requirements for {\em pure} photoionization of the web are 
nevertheless hard to reconcile with the observed H$\alpha$ fluxes. 
A combination of shock- and photoionization could therefore pose a viable
explanation for the H$\alpha$ emission of the web, for example in a
scenario where the web is a pre-existing neutral halo cloud that has been 
moderately shocked by the wind -- perhaps somewhat similar to, but brighter 
than, the X-ray/H$\alpha$ ridge seen well above the disc of M82 
\citep*{lehn1999}.

\subsection{Fate of the diffuse gas}\label{sec,fate}

\citet{hunt1996} conjectures that if the H$\alpha$ features represent 
fragmenting supershells, it is likely they will break out of the galaxy since
they have already reached a large distance (1--2~kpc) from the midplane.
For the X-ray gas, the derived cooling time is sufficiently large 
that radiative losses 
can be neglected as a first approximation. In that case, gas with an 
adiabatic index of $\gamma=5/3$ hotter than
\begin{equation}
T_{esc} = 1.1\times 10^5 (v_{esc}/100 \mbox{ km s$^{-1}$})^2 \mbox{ K}
\label{eq,t_esc}
\end{equation}
should be able to escape the galactic gravitational potential 
\citep{wang1995}.
Modelling the galactic potential as a singular isothermal sphere,
the local escape velocity $v_{esc}$ at a distance $r$ from the centre is 
given by
\begin{equation}
v_{esc}(r) = \sqrt{2}v_{max}[1+\mbox{ln}(r_t/r)]^{1/2},
\end{equation}
where $v_{max}$ is the maximum rotation velocity, and the potential is
truncated at $r=r_t$. 
With H{\sc i} having been detected out to 3~times the adopted major radius 
of $D_{25}$ \citep{hunt1994},
i.e.\ $\sim 7$~kpc, we can conservatively choose $r_t=50$~kpc. 
H{\sc i} observations further suggest 
$v_{max} \approx 30$~km~s$^{-1}$ (LEDA database), yielding 
$v_{esc}\sim 100$~km~s$^{-1}$, and hence $T_{esc} \sim 0.01$~keV, 
for the X-ray gas at either tip of the X-ray 
emission ellipse $D_{spec}$ ($r \approx 2$~kpc). This is well below the
derived gas temperature of $T\simeq 0.25$~keV and suggests that a 
blow-out of gas from the galaxy is a possibility. 

In practice, however, 
the extended H{\sc i} halo may prevent blow-out from occurring.
Following Summers et~al.\ (2003, 2004), we can evaluate this possibility
using the blow-out criterion of \citet{macl1988}. Again neglecting radiative 
losses, this criterion predicts wind blow-out in a disc galaxy provided that 
the dimensionless quantity 
\begin{equation}
\zeta = 9.4L_{mech,40} H_{z,kpc}^{-2} P_4^{-3/2} n_0^{1/2}
\label{eq,escape}
\end{equation}
exceeds unity. Here $L_{mech,40}$ is the mechanical energy luminosity of
stellar winds and supernova ejecta in
units of $10^{40}$~erg~s$^{-1}$, $H_{z,kpc}$ is the galactic vertical 
scaleheight in kpc, $P_4$ is the pressure of the surrounding ISM in units of 
$P/k = 10^4$~K~cm$^{-3}$, and
$n_0$ is the ISM number density in cm$^{-3}$ at the disc midplane. 
In the case of complete thermalization of the mechanical energy output from 
supernovae, $L_{mech}$ corresponds to the 
rate at which the wind thermal energy content $E_{th}$ 
(Table~\ref{tab,phys}) has been 
injected over the lifetime of the starburst. Having down-scaled the outflow
age to $\sim 50$~Myr instead of using the $\eta=1$ expectation of 
$\sim 100$~Myr, we should reduce $E_{th}$ accordingly. This would suggest 
$L_{mech,40} \approx 0.2$, independently of $\eta$. The observed scaleheight 
can be deduced from the H{\sc i} profile derived by \citet{hunt1994},
suggesting $H_z\approx 2.5$~kpc. 
For an exponential atmosphere,
the column density $N_{\mbox{\footnotesize H{\sc i}}} \sim H_z n_0$, 
implying $n_0 \approx 0.1$~cm$^{-3}$ for 
$N_{\mbox{\footnotesize H{\sc i}}} \sim 10^{21}$~cm$^{-2}$ as 
suggested by the H{\sc i}~map of \citet{hunt1994}. Conservatively assuming 
a characteristic ISM temperature of $10^4$~K, 
i.e.\ $P_4 \approx 0.1$, we then have $\zeta \approx 3$, and the blow-out 
condition is easily satisfied. While only an indicative result, this 
clearly suggests as a first approximation that the halo is unable to retain 
the wind. Moreover, a criterion similar to equation~(\ref{eq,escape}), but 
slightly less strict, is adopted by \citet{koo1992}
(the constant in equation~[\ref{eq,escape}] effectively being $\approx 15$ 
rather than 9.4), which also suggests 
that for a realistic choice of parameters, the wind may indeed escape
NGC\,1800 and its H{\sc i}~halo. 
Any patchiness of the H{\sc i}~halo may further facilitate blow-out 
along certain directions through the halo. 

The question remains whether the ICM, rather than the galactic halo, can
confine the wind. Limited work has been done in addressing the impact of 
external pressure on the possibility of blow-out. One example is the
model of \cite{sili2001} in which ICM pressure is taken into account
through the way it modifies the ISM distribution, assuming pressure balance 
at the galaxy--ICM boundary. At the relatively low ICM pressures considered
in that study ($P_{ICM}/k=1$ and 100~K~cm$^{-3}$), it is the mass of any
extended gaseous halo, rather than $P_{ICM}$, which sets the effective minimum
value of $L_{mech}$ required for blow-out. For their specific models, a 
disc-like galaxy with $M_{ISM}\sim 10^8$~M$_\odot$ and 
$M_{halo}=10^8$~M$_\odot$ (presumably reasonably representative of NGC\,1800),
embedded in an ICM with $P_{ICM}/k=100$~K~cm$^{-3}$, blow-out demands 
$L_{mech,40}\ga 0.1$. This result is well in line with our estimate based on 
equation~(\ref{eq,escape}), suggesting that the wind should be able to expand
freely into the ambient ICM. It is not clear, however, how to extend the model
results of \cite{sili2001} to the exact situation in NGC\,1800, given the
more subtle dependence on the -- in this case unknown -- dark matter mass of
the galaxy. Moreover, it is also unclear how to extrapolate the results to
larger $P_{ICM}$ and how to deal with effects of additional ram-pressure due 
to the
motion of the galaxy through the ICM. Such effects are
included in the hydrodynamical simulations of \cite{mura1999}. Initially
neglecting ram-pressure, these authors find, for a fiducial model dwarf of 
dark matter mass $M=10^9$~M$_\odot$, that a starburst wind confined by 
ICM thermal pressure reaches a maximum radius of expansion $R_{max}$ which, 
for $R_{max} \la 2$~kpc (cf.\ Fig.~\ref{fig,image2}), would correspond to 
$P_{ICM}/k \ga 5\times 10^4$~K~cm$^{-3}$. For the ICM pressure, our revised 
group luminosity estimate of 
$L_X<10^{41}$~erg~s$^{-1}$ would imply $T\la 0.4$~keV, assuming the
$L_X$--$T$ relation of \citet{hels2000}. For a $Z=0.2$Z$_\odot$ $T=0.4$~keV
{\sc mekal} plasma, our constraint on $L_X$ translates, 
via equation~(\ref{eq,xspec}), into a maximum central ICM density of 
$N_0\la 8\times 10^{-4}$~cm$^{-3}$, assuming a group radius of 300~kpc and, 
conservatively, $\beta$-model parameters of $r_c=50$~kpc and $\beta=0.5$ 
(cf.\ Section~\ref{sec,group}). The resulting ICM thermal pressure, 
$P/k < 4\times 10^3$~K~cm$^{-3}$, is at least an
order of magnitude below that required for confinement of the wind at its
current height above the plane. As for ram-pressure, the radial
velocity $v$ of NGC\,1800 with respect to the group centre is 
$\sim 400$~km~s$^{-1}$ \citep{maia1989}, suggesting $P_{ram} =
\rho v^2 \la 1.3 \times 10^{-12}$~dyn~cm$^{-2}$ ($\sim 2.5 P_{ICM}$) and 
hence a total pressure of $P_{tot}/k \la 1.4\times 10^4$~K~cm$^{-3}$. 
This does not change the conclusion that the outflow probably remains 
unimpeded by the ICM at present; indeed, the 
simulations of \cite{mura1999} show that for $P_{ICM}/k=10^3$~K~cm$^{-3}$ and
$v=600$~km~s$^{-1}$, a wind is only marginally affected during its 
expansion phase relative to the case of $P_{ram}=0$.

As a final test, we can obtain a rough idea about the extent to which the
wind is actually confined, by considering the surface brightness structure 
$S(r)$ of the outflow. A freely expanding, spherically symmetric wind is 
expected to show a power-law decline in volume density of the wind fluid,
$n \propto r^{-\alpha}$, with $\alpha=2$ and where $r$ is the distance 
along the outflow \citep{chev1985}. A similar
behaviour is expected for a conical outflow of constant opening angle, 
which is perhaps more relevant for NGC\,1800. A power-law fit, 
$S \propto r^{-\gamma}$, to the 
0.3--2 keV profile shown in Fig.~\ref{fig,surfbright2} gives 
$\gamma \approx 1.3\pm 0.1$ for both the northern and southern ends of a 
conical outflow.
Since, at fixed cone height $r$, the line of sight through the cone grows 
with $r$, we have $S(r)\propto n^2 r$ for an isothermal wind expanding in a 
regular cone, and hence $\alpha = (\gamma+1)/2 \approx 1.15$.  
Temperature variations within the wind would have limited impact on this 
result, since $S$ is roughly independent of $T$ at the relevant temperatures
($\sim 0.25$ keV) and energies (0.3--2 keV).
The freely expanding winds of M82 and NGC\,253 show values of 
$\alpha = 0.9$ and 1.3, respectively, thus bracketing our estimate 
of $\alpha \approx 1.15$.
By contrast, NGC\,3077, in which X-ray and H$\alpha$ images suggest that 
the hot X-ray bubbles are indeed confined by H$\alpha$ gas, shows a value 
$\alpha \approx 0.6$ (see \citealt{ott2003} for a discussion of the 
results for these three galaxies). Overall, this seems to indicate that the
wind of NGC\,1800 is not substantially confined. 

Summarizing, there is no indication
of significant confinement from the surface brightness structure of the wind,
a result clearly supported both by models and simulations of dwarf
galaxy winds expanding into an ambient medium of the relevant pressures.
The above considerations suggest that the wind, currently
appearing to be breaking out of the galactic plane, will eventually
blow out through the galactic halo, 
probably delivering energy and newly synthesized metals to the ICM. 
This would be particularly true if the very hot wind fluid,
presumably remaining undetected in these observations but carrying most of 
the wind energy and metals (cf.\ \citealt{stri2000b}), can have terminal
outflow velocities much larger than 
those of the X-ray detectable gas (e.g.\ \citealt{heck2003}).

\section{Scaling properties of dwarf starburst galaxies}\label{sec,comparison}

Due to their X-ray faintness, detailed X-ray studies of dwarf galaxies have 
generally not been feasible prior to the advent of {\it Chandra} and
{\it XMM-Newton}. Consequently,
the small number of dwarf starbursts studied in X-rays so far has precluded
any statistical investigation of their X-ray properties, although 
\citet{hart2004} discussed some properties of the three dwarf starbursts 
NGC\,4214, NGC\,4449, and NGC\,5253. To amend this
situation, the derived properties of NGC\,1800 are here compared to those of
other dwarf starbursts and to starburst galaxies in general, and a first 
detailed discussion is presented of the X-ray properties of all dwarf 
starburst galaxies that have so far been subjected to detailed X-ray studies
and in which, to our knowledge, diffuse X-ray emission has been unambiguously
detected.

In Table~\ref{tab,dwarfs} we list some properties of these dwarf starbursts 
and the associated references to published X-ray data. 
\begin{table*}
\centering
 \begin{minipage}{149mm}
\caption{\protect{Summary of properties of dwarf starbursts. 
X-ray luminosities are total values for the diffuse gas only.}}
\label{tab,dwarfs}
\begin{tabular} {lcrrrrl} \hline 
\multicolumn{1}{l}{Galaxy} &
\multicolumn{1}{c}{$D$} &
\multicolumn{1}{c}{$L_X$} &
\multicolumn{1}{c}{$L_{FIR}$} & 
\multicolumn{1}{c}{$L_B$} & 
\multicolumn{1}{c}{$L_K$} & 
\multicolumn{1}{c}{References} \\  
 & (Mpc) & ($10^{38}$ erg s$^{-1}$) & \hspace{3mm}($10^8$ L$_{\odot}$) & ($10^8$ L$_{\odot}$)\hspace{4mm} & ($10^8$ L$_{\odot}$)  & \\ \hline
NGC\,625  & 3.9 & $\sim 1.0^a$ \hspace{4mm}   & \hspace{2.5mm} $1.3\pm0.1^i$     & 5.4\hspace{4mm}   & $9.0\pm0.4$\hspace{0mm} & \citet{boma1998}\\
NGC\,1569 & 2.2 & $8.2^b$ \hspace{4mm}      & \hspace{2.5mm} $3.5\pm0.4^i$     & $10.4\pm0.6$\hspace{4mm}   & $7.5\pm0.2$\hspace{0mm}  & \citet{mart2002}\\
NGC\,3077 & 3.6 & 20--50$^c$ \hspace{4mm}   & \hspace{2.5mm} $1.4\pm0.1^j$ & $12.8\pm1.0$\hspace{4mm}   & $33.6\pm0.6$\hspace{0mm}  & \citet{ott2003}\\
NGC\,4214 & 2.9 & $3.5\pm 0.3^d$ \hspace{4mm}       & \hspace{2.5mm} $2.4\pm0.3^i$     & $11.7\pm0.9$\hspace{4mm}  & $12.7\pm0.6$\hspace{0mm} & \citet{hart2004}\\
NGC\,4449 & 2.9 & $9.1\pm 0.6^d$ \hspace{4mm}      & \hspace{2.5mm} $9.5^k$     & $17.4\pm1.0$\hspace{4mm}  & $23.2\pm0.8$\hspace{0mm} & \citet{summ2003}\\
NGC\,5253 & 3.2 & $3.9\pm 0.3^e$ \hspace{4mm}      & \hspace{2.5mm} $3.9\pm0.4^i$     & $7.0\pm0.6$\hspace{4mm}   & $10.3\pm0.3$\hspace{0mm}  & \citet{summ2004}\\
NGC\,1800 & 7.4 & $1.3\pm 0.3^f$ \hspace{4mm}      & \hspace{2.5mm} $0.8\pm0.1^i$     & $5.0\pm0.4$\hspace{4mm}   & $9.7\pm0.5$\hspace{0mm}  & This study\\
\hline \\
\end{tabular}
Notes: $^a$0.1--2.4 keV. Uncertainties on $L_{\rm X}$ are not available for 
this galaxy. $^b$0.3--6 keV; uncertainties on $L_{\rm X}$ are 
not available. $^c$$L_X$ varies from $20-50\times 10^{38}$ erg s$^{-1}$ 
depending on the spectral model assumed. $^d$0.3--8 keV. $^e$0.37--6.0 keV. 
This galaxy also displays spectral evidence for a diffuse power-law 
component, possibly due to unresolved point sources; the quoted value is for 
the thermal components alone. $^f$0.3--3 keV.  
$^i$40--120 $\mu$m, see text for details. $^j$40--120 $\mu$m; 
\protect{\citet{yun1999}} lists a value $2.7\times 10^8$ L$_{\odot}$ in an 
unspecified band. $^k$10--150 $\mu$m \citep{summ2003}, see text.
\end{minipage}
\end{table*}
We have assumed distances $D$ as given in the reference in the Table, 
and $B$-band 
luminosities from \citet{tull1988}, corrected to these distances, with
uncertainties based on those of the apparent magnitudes as listed by
\citet*{deva1976} (for NGC\,625, no values are provided). Note that 
NGC\,1800 is the most distant dwarf starburst having diffuse X-ray emission
safely detected so far (\citealt{thua2004} report marginal {\em Chandra}
detections of diffuse emission in the $D=54.3$ Mpc and $D=12.6$ Mpc dwarfs 
SBS\,0335-052 and I\,Zw\,18, but at a level of just 
$8\pm 5$ and $23\pm 7$ diffuse source counts, respectively).
$K_s$ band luminosities in the Table have been computed for $D$ from the 
total $K_s$ magnitude
of the galaxy as listed in the 2MASS Extended Sources Catalog\footnote{http://irsa.ipac.caltech.edu/applications/2MASS/PubGalPS/}. 
The far-infrared (40--120 $\mu$m) luminosities
$L_{FIR}$ have been calculated from the 60 $\mu$m and 100 $\mu$m fluxes 
$S_{60}$ and $S_{100}$ and their errors as listed in the 
IRAS Galaxy Catalog \citep{full1989}, using the relation of \citet{deve1989}:
\begin{equation}\label{eq,lfir}
L_{FIR} = 3.65 \times 10^5 [2.58S_{60}\mbox{(Jy)} + S_{100}\mbox{(Jy)}] D^2 \mbox{(Mpc)} \mbox{ L}_{\odot}.
\end{equation}
An exception is NGC\,4449 which is not listed in the IRAS Galaxy Catalog. 
Instead we adopt the 10--150 $\mu$m value also used by \citet{summ2003}.
We have chosen not to attempt to correct this value to the
40--120 $\mu$m band, as the typical FIR spectral energy distribution of 
local starbursts rises sharply between $\sim 10$--100 $\mu$m and then 
declines, making the difference between the two bands relatively small (e.g.\ 
\citealt{spin2002}).

The reason for tabulating $L_{FIR}$, $L_B$, and $L_K$ along with the X-ray 
luminosity of the diffuse gas, is that $L_{FIR}$ is known
to be a tracer of the star formation rate (Section~\ref{sec,hot}) and
as such might be related to the diffuse X-ray luminosity from the ISM and
a starburst wind. Also, $L_K$ should provide a reliable measure of stellar 
mass, but for comparison with \citet{read2001} we also examine dependencies 
on $L_B$, although this quantity is more affected by the presence of young
stellar populations and dust.
Following \citet{read2001}, we can then define the 'activity' of
the galaxy as the star formation rate per unit stellar mass, measured by 
$L_{FIR}/L_B$, or even better, $L_{FIR}/L_K$.

Inspired by the multivariate statistical study of 
\citet[hereafter RP01]{read2001} 
on normal (i.e.\ low-activity) dwarfs and non-dwarf starbursts, 
we have performed the first investigation of the X-ray scaling properties
of dwarf starbursts with well-studied diffuse X-ray emission. 
To bring all galaxies on an equal footing and obtain consistency with RP01, 
we have converted dwarf X-ray luminosities to the
0.1--2 keV band, assuming a {\sc mekal} plasma with $Z=0.2$Z$_\odot$ at the
relevant temperature.
Linear regression fits for the parameter combinations
under study (in logarithmic space) have been conducted using different 
regression methods \citep{isob1990}. Results for orthogonal 
regression fits are presented in Table~\ref{tab,regress}; only those
parameter combinations yielding acceptable fits 
are reproduced in the Table, taken to be those fits for
which the results of different regression techniques 
-- orthogonal regression, ordinary least-squares (OLS), OLS bisector, 
mean OLS, and reduced major-axis -- gave consistent results.
Also listed in the Table is the value of the Spearman 
rank-order correlation coefficient $r_s$ (in the range $[-1;1]$) and the 
significance $t_s$ of $r_s$ being non-zero (see e.g.\ \citealt{pres1996}).

\subsection{X-ray vs optical and far-infrared luminosities}
We plot in Fig.~\ref{fig,lxlfir}a a comparison of
the far-infrared and diffuse X-ray luminosities of these dwarf 
starbursts. Also plotted for comparison are the eight non-dwarf starbursts 
in the sample of RP01, defined by these authors as those galaxies 
having $L_{FIR}>0.38L_B$ (with $L_{FIR}$ also derived via 
equation~[\ref{eq,lfir}]).
To be able to compare our results directly to those of 
RP01, we use $L_B$ as the third parameter of investigation 
but compare the results to those obtained using $L_K$ below.
There is, in fact, a tight correlation between $L_B$ and $L_K$ for 
{\em all} the galaxies shown in the figure ($r_s=0.93$, $t_s=9.0$), 
suggesting that $L_B$ may indeed be used as a proxy for the stellar mass of 
these galaxies.
\begin{figure*}
\begin{center}
\mbox{\hspace{-0.5cm}
\epsfxsize=9.2cm
\epsfysize=7.4cm
\epsfbox{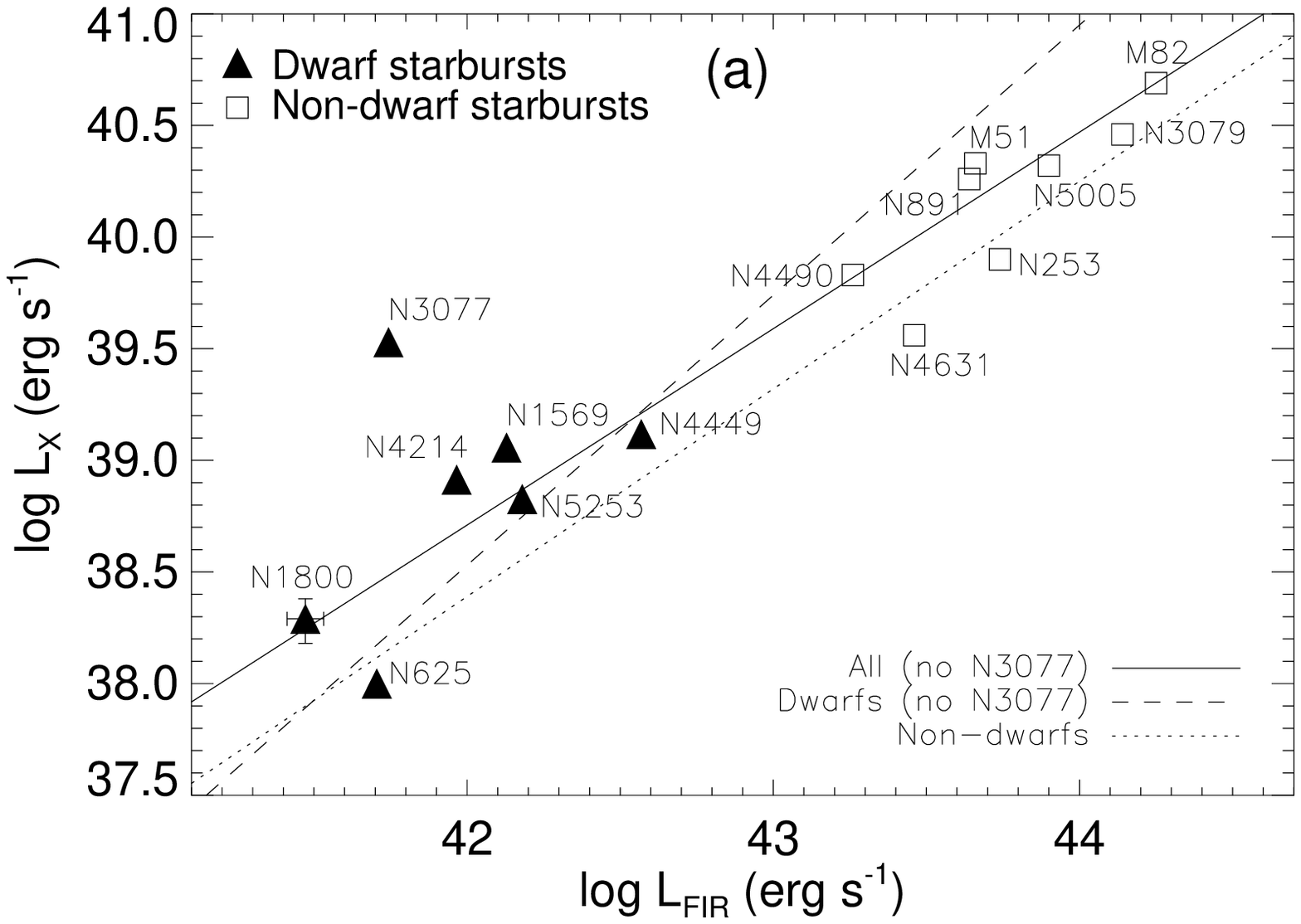}\hspace{0cm}
\epsfxsize=9.2cm
\epsfysize=7.4cm
\epsfbox{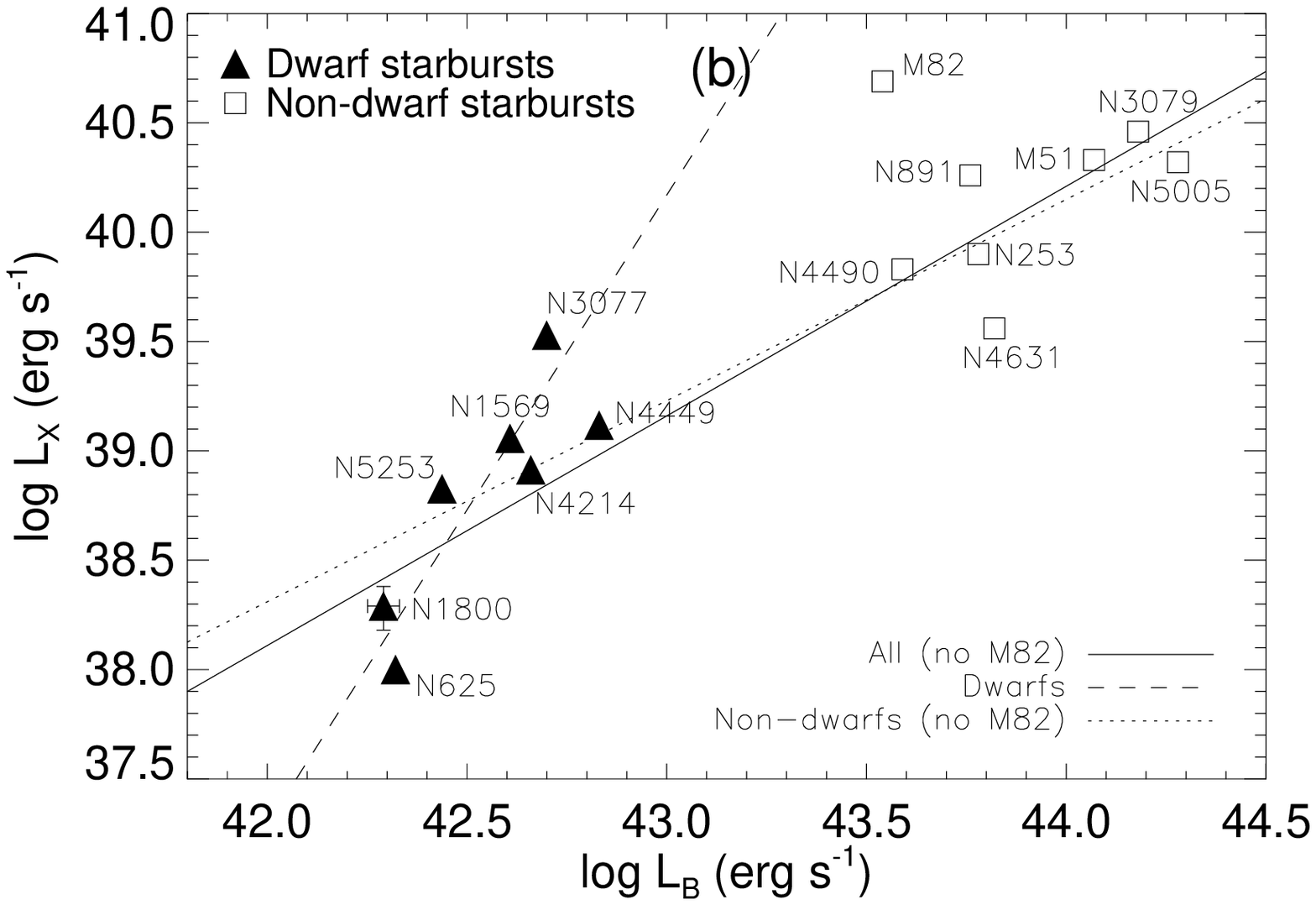}\hspace{0cm}}
\caption{Diffuse $L_X$ along with far-infrared luminosity $L_{FIR}$ (a), and 
$B$-band luminosity $L_B$ (b) for starburst galaxies. Dotted lines are the 
best-fitting relations of Read \& Ponman (2001) to {\it ROSAT} 0.1--2 keV 
data of a subsample of eight starburst spirals, 
log $L_X = (0.93\pm0.26)$ log $L_{FIR} - (0.67\pm0.22)$, and (excluding M82) 
log $L_X = (0.92\pm0.42)$ log $L_B - (0.33\pm0.26)$. Dashed lines represent
the relations obtained from our regression analyses of the dwarf subsample
(see Table~\ref{tab,regress}), and solid lines those derived for the full
(dwarf + non-dwarf) sample. For clarity in these and the following figures, 
typical error bars for the dwarf galaxies (cf.\ Table~\ref{tab,dwarfs}) are 
shown for NGC\,1800 only.}
\label{fig,lxlfir}
\end{center}
\end{figure*}
With one notable exception (NGC\,3077), $L_X$ is seen to correlate  
with $L_{FIR}$ for the dwarfs, as was found by RP01 for non-dwarfs and 
quiescent galaxies. The dwarf correlation seems to match well that derived 
for such galaxies,
leading to a remarkably tight correlation ($t_s=11.3$) across three
orders of magnitude in $L_{FIR}$ for the combined sample of dwarf and normal
starburst galaxies. NGC\,1800 lies at the low end of the range in $L_X$ and 
$L_{FIR}$ spanned by the dwarf starbursts, but is otherwise fairly typical, 
fitting perfectly well to the correlation derived for the full starburst 
sample. Taken as a whole, $L_X$ and $L_{FIR}$ for the dwarf starbursts
are well matched to values for normal non-dwarf spirals but lie substantially
below those of non-dwarf starbursts. These results are to be 
expected, if the observed diffuse X-ray emission is indeed associated with 
the global level of star formation.
In Fig.~\ref{fig,lxlfir}b we plot correspondingly $L_X$
and $L_B$. There is a clear correlation between these two parameters
for the dwarf starbursts ($t_s=4.4$). 
This correlation appears to be somewhat steeper 
than is the case for non-dwarf starbursts, but not highly significantly so if 
NGC\,3077 is excluded from the dwarf sample (the regression slopes then being 
consistent at the $\sim2\sigma$ level). 
The reasonable match between the dwarf and non-dwarf correlations again lead
to a tight correlation for the combined sample ($t_s = 9.8$, with M82 
excluded) across two orders of magnitude in $L_B$.
The correlations are less significant
when using $L_K$ instead of $L_B$, with $t_s=1.9$ for dwarfs, 2.2 for 
non-dwarfs, and 8.1 for the combined sample. The dwarf regression slope 
$m=2.85\pm 0.96$ is also in this case steeper than that of normal starbursts, 
($m=0.98\pm0.40$), but again consistent with it at the $\sim2\sigma$ level.

In Fig.~\ref{fig,lxlfir2}a the quantity measuring galaxy activity 
($L_{FIR}/L_K$) is plotted against stellar mass ($L_K$).
No clear correlation is seen, with $t_s = -0.9$ for dwarfs and $-0.4$ for the 
entire sample, giving unacceptable regression fits. This is analogous to the 
result for the larger and more general sample of RP01 based on 
$B$-band luminosities. It implies that we may separate the impact 
of mass and activity on our results.
\begin{figure*}
\begin{center}
\mbox{\hspace{-0.5cm}
\epsfxsize=9.2cm
\epsfysize=7.4cm
\epsfbox{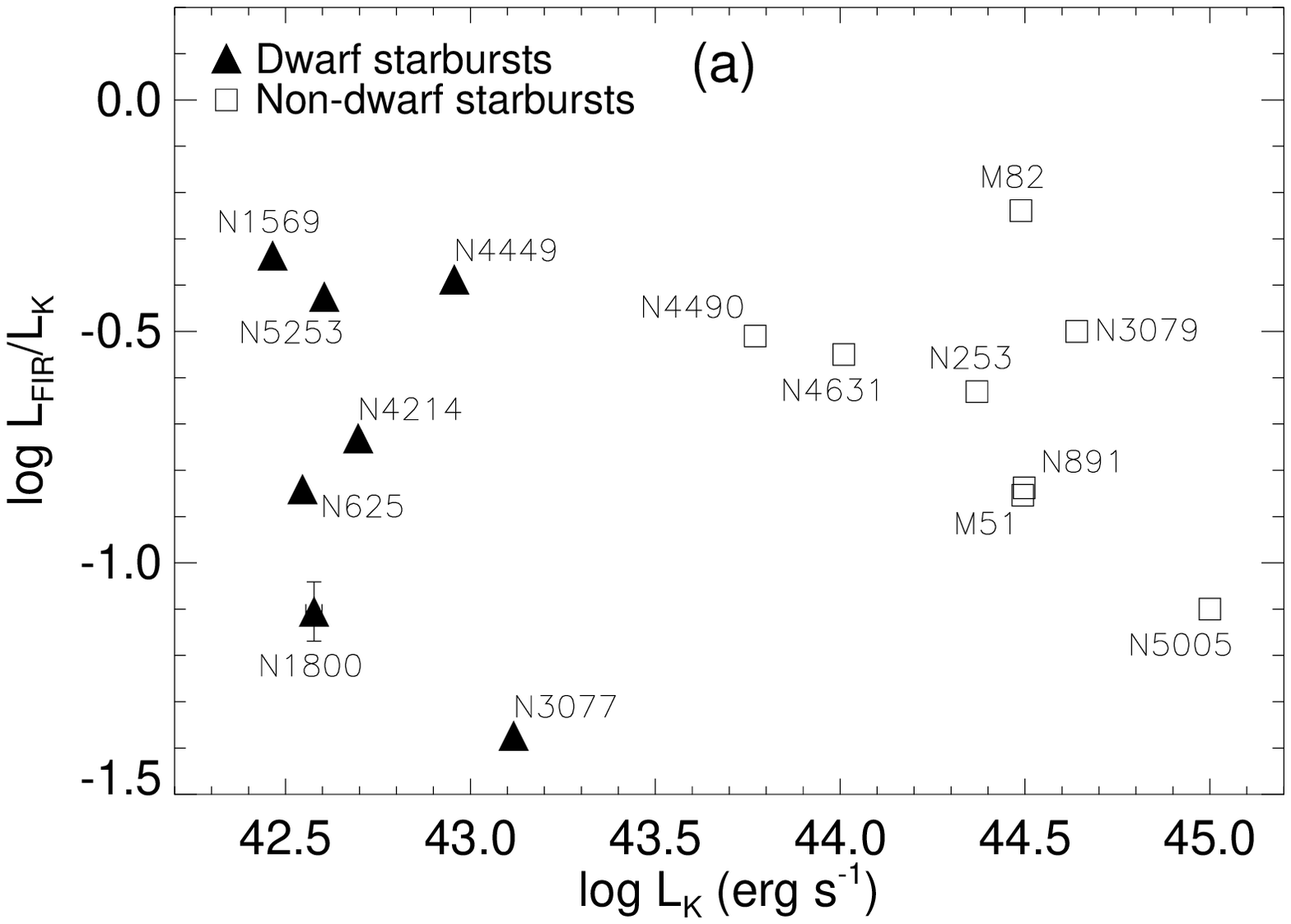}\hspace{-0.0cm}
\epsfxsize=9.2cm
\epsfysize=7.4cm
\epsfbox{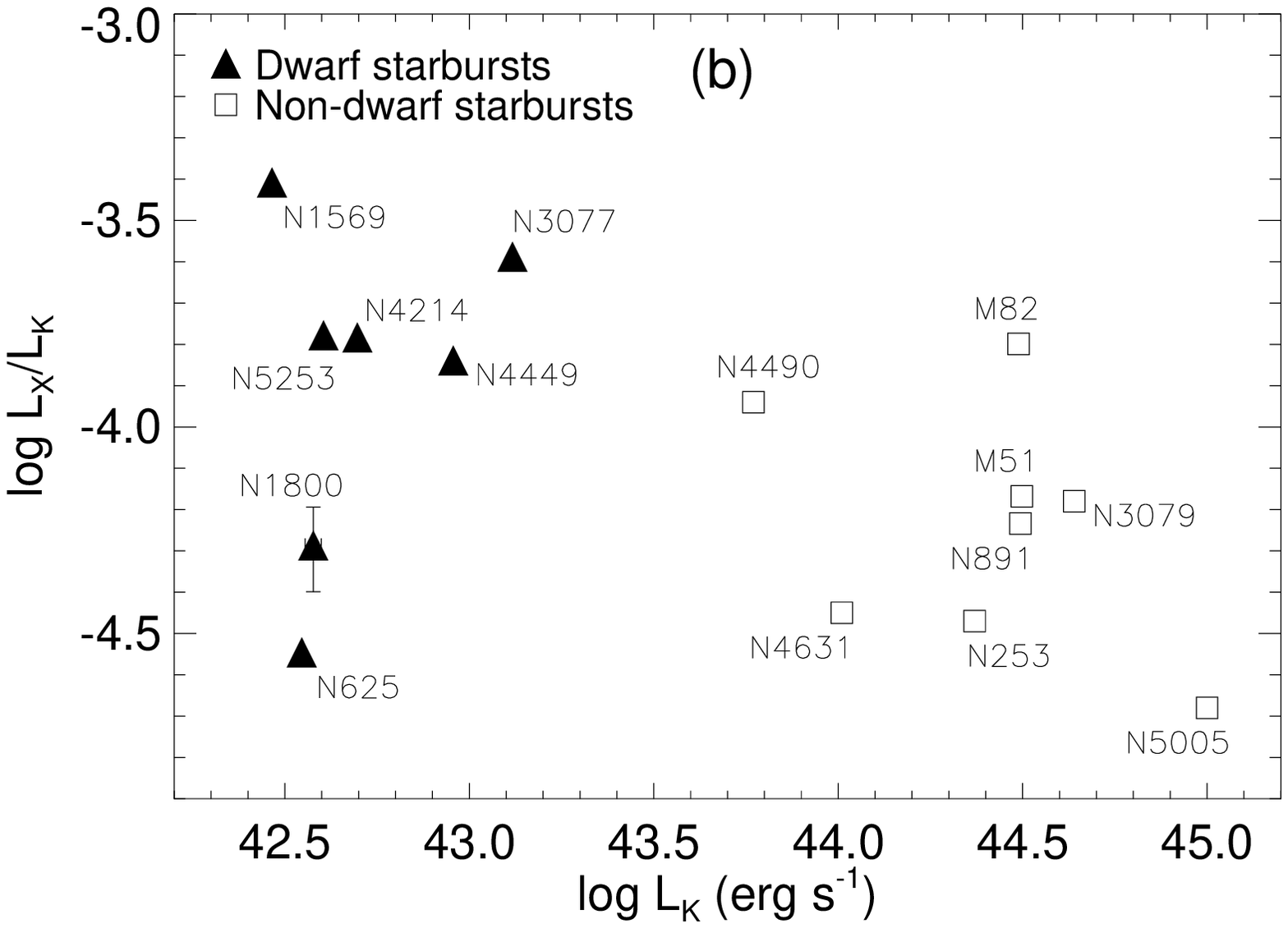}\hspace{0cm}}
\caption{Star formation 'activity' $L_{FIR}/L_K$ (a) and 'mass-normalized' 
X-ray luminosity $L_X/L_K$ (b) as a function of galactic stellar 'mass' 
$L_K$.}
\label{fig,lxlfir2}
\end{center}
\end{figure*}
As can be seen, the activity of NGC\,1800 lies at the low end of 
typical starbursts (cf.\ also Fig.~\ref{fig,lxlfir}a).
The 'mass-normalized' X-ray luminosity $L_X/L_K$ as plotted in 
Fig.~\ref{fig,lxlfir2}b does not show a clear
correlation with $L_K$ either ($t_s=0.2$ for dwarfs, $-0.7$ for normal 
starbursts, and $-1.5$ for the full sample, producing unacceptable fits in all
cases). This is again consistent with the results
of RP01 for more massive starbursts, using $L_B$ rather than 
$L_K$, whereas the 'normal', i.e.\ quiescent, galaxies of that study display 
a strong correlation ($t_s=4.7$). 
The position of NGC\,1800 in Fig.~\ref{fig,lxlfir2}b is 
not conspicuous and demonstrates no evidence for enhanced 
wind luminosity due to compression of the wind fluid by ICM confinement. 
NGC\,1569 and NGC\,3077, on the other hand, appear to have a somewhat
higher $L_X/L_K$ ratio than the remaining starbursts.

When plotted against the star formation 'activity' $L_{FIR}/L_K$ in 
Fig.~\ref{fig,lxlfir3}, $L_X/L_K$ of the dwarf
starbursts does seem to show a dependence, albeit a weak one,
and again with NGC\,3077 clearly deviating from the common trend 
(this galaxy must be excluded in order to obtain an acceptable fit for the 
dwarfs alone). Aside from NGC\,3077, the dwarf starbursts 
conform to the relation found for the starburst sample of RP01.
Again, NGC\,1800 seems to be fairly typical of the dwarf starburst population,
which, in terms of absolute numbers, is similar to non-dwarf starbursts in
this representation but shows slightly higher values on both axes than
do quiescent spirals. 
If using $L_B$ rather than $L_K$ as a
tracer of stellar mass, the correlations between activity and 
mass-normalized $L_X$ become less significant ($t_s = 1.7$ for dwarfs and
2.1 for the combined sample), with comparable regression slopes for dwarfs 
and normal starbursts, as seen from Table~\ref{tab,regress} and from the 
relation derived by RP01 for their sample, 
log $L_X/L_B$ = $(0.96\pm0.23)$ log $L_{FIR}/L_B - (3.59\pm0.23)$.
Interestingly, it also appears that dwarf starbursts generally have lower 
values of $L_{FIR}/L_B$ (in the range $[-1.0;-0.25]$) 
compared to normal starbursts ($[-0.4;0.7]$). 
As can be seen from Fig.~\ref{fig,lxlfir3}, there is no such 
indication for $L_{FIR}/L_K$, suggesting that the effects of relatively 
stronger dust obscuration in larger galaxies is responsible. In addition,
dwarf starbursts are low-metallicity objects due, at least in part, to their
shallow gravitational potentials. Assuming a dust-to-gas mass ratio which is 
proportional to metallicity (\citealt{lise1998}; \citealt{gall2003}), dwarf
starbursts would thus on average suffer considerably less obscuration than 
larger starbursts, as observed.

NGC\,3077 seems to deviate somewhat from the common trend depicted in 
Figs.~\ref{fig,lxlfir}a and \ref{fig,lxlfir3}.
This galaxy is a
member of a group/triplet consisting also of M81 and the prototypical 
starburst M82. There is substantial H{\sc i} evidence that these three 
galaxies are tidally interacting,
and these interactions are believed to have triggered the starburst activity 
in both NGC\,3077 and M82 (see e.g.\ \citealt{ott2003}). It is  
suggested by
Figs.~\ref{fig,lxlfir}--\ref{fig,lxlfir3} that NGC\,3077 exhibits a
current star formation activity which is at the low end of typical dwarf 
starbursts, while at the same time displaying a rather high X-ray 
luminosity. This seems particularly true in
light of the fact that we have plotted the {\em lowest} value of $L_X$ 
allowed by the analysis of \citet{ott2003}. Oddly, this result is exactly 
opposite to that derived by \citet{read1998} on high-activity merging and
interacting systems, in which a {\em deficit} of X-ray emission relative to
$L_{FIR}$ was found in the most active systems. 
The physical scale of soft diffuse emission in NGC\,3077 is less than 1 kpc, 
but the {\em Chandra} observations of \citet{ott2003} show no evidence of 
strong nuclear non-stellar activity. Any AGN would remain largely hidden in 
X-rays, since only $\sim 15$ per cent of all X-ray emission within this
galaxy originates in resolved point sources. A hidden AGN would 
presumably be dust-enshrouded and so should radiate strongly in the FIR, for
which there is no evidence in Figs.~\ref{fig,lxlfir}--\ref{fig,lxlfir3}.
It is possible that the star formation rate of NGC\,3077 has recently 
declined, and that we are currently seeing hot gas generated from star 
formation which took place a few Myr ago. This is, in fact, in line with the 
results of \citet{ott2003}
who find that the hot bubble properties in NGC\,3077 suggest a 
star formation rate of $\sim 0.6$ M$_{\odot}$ yr$^{-1}$, an order of 
magnitude above the current SFR of $0.06$ M$_{\odot}$ yr$^{-1}$ as derived 
from H$\alpha$ and FIR fluxes (note also that a rise in $L_{FIR}$ by a factor
of 
$\sim 10$ would actually drop NGC\,3077 on to the general trends shown in 
Figs.~\ref{fig,lxlfir}a and \ref{fig,lxlfir3}). With an estimated bubble age 
in NGC\,3077 of 
2--10 Myr, this may indicate that the SFR of this galaxy has declined 
considerably within the last few Myr, and that the observed X-ray luminosity
to a large degree reflects previous rather than current star formation 
activity.  
\begin{figure}
\begin{center}
\mbox{\hspace{-0.6cm}
\epsfxsize=9.3cm
\epsfysize=7.4cm
\epsfbox{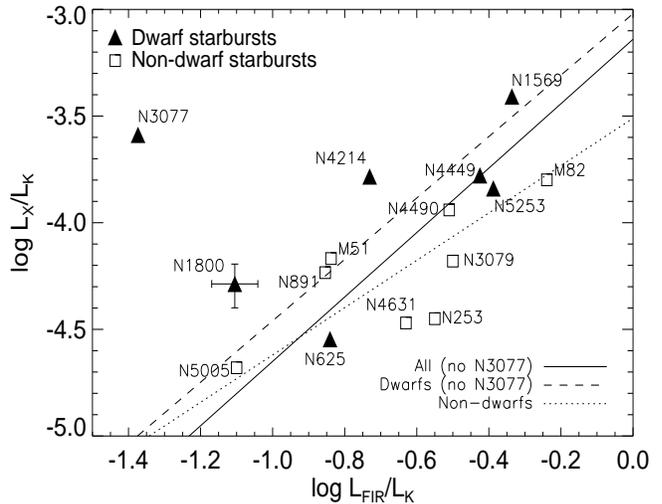}}
\caption{Diffuse $L_X$ and far-infrared luminosity $L_{FIR}$ for starbursts,
'mass-normalized' using $L_K$. Lines have the same meaning as for 
Fig.~\ref{fig,lxlfir}a.}
\label{fig,lxlfir3}
\end{center}
\end{figure}

\subsection{Dependencies on gas temperature}
Motivated by the presence of scaling relations between $T$ of X-ray gas 
and e.g.\ $L_X$ and mass derived for galaxy groups and clusters, we plot in 
Fig.~\ref{fig,lxlfir4} the diffuse X-ray luminosities, hot gas masses, and
galactic stellar 'masses' as functions of the hot gas temperature $T$. 
\begin{figure}
\begin{center}
\vspace{-0.25cm}
\mbox{\hspace{-0.5cm}
\epsfxsize=9.2cm
\epsfysize=6.8cm
\epsfbox{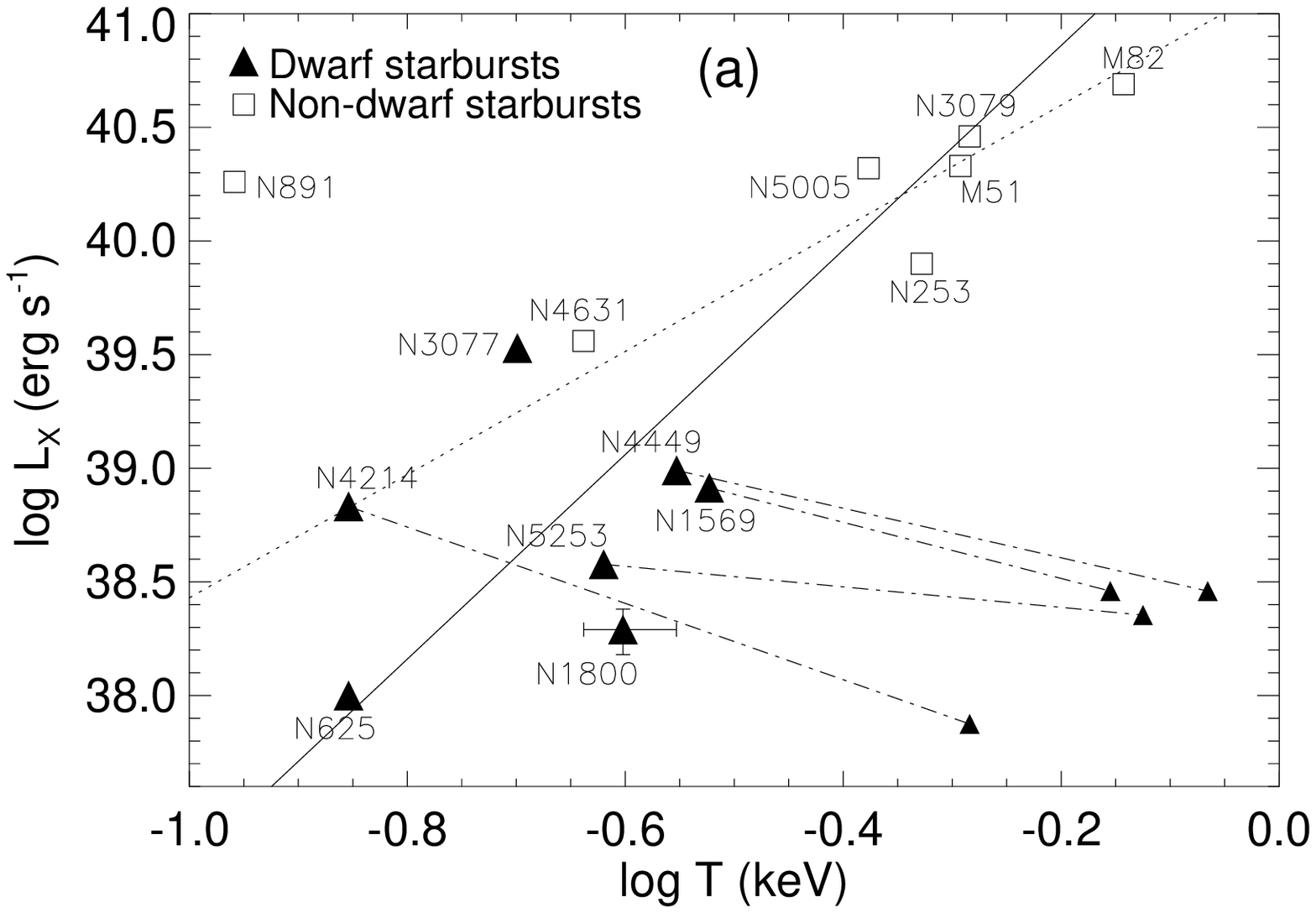}}
\mbox{\hspace{-0.5cm}
\epsfxsize=9.2cm
\epsfysize=6.8cm
\epsfbox{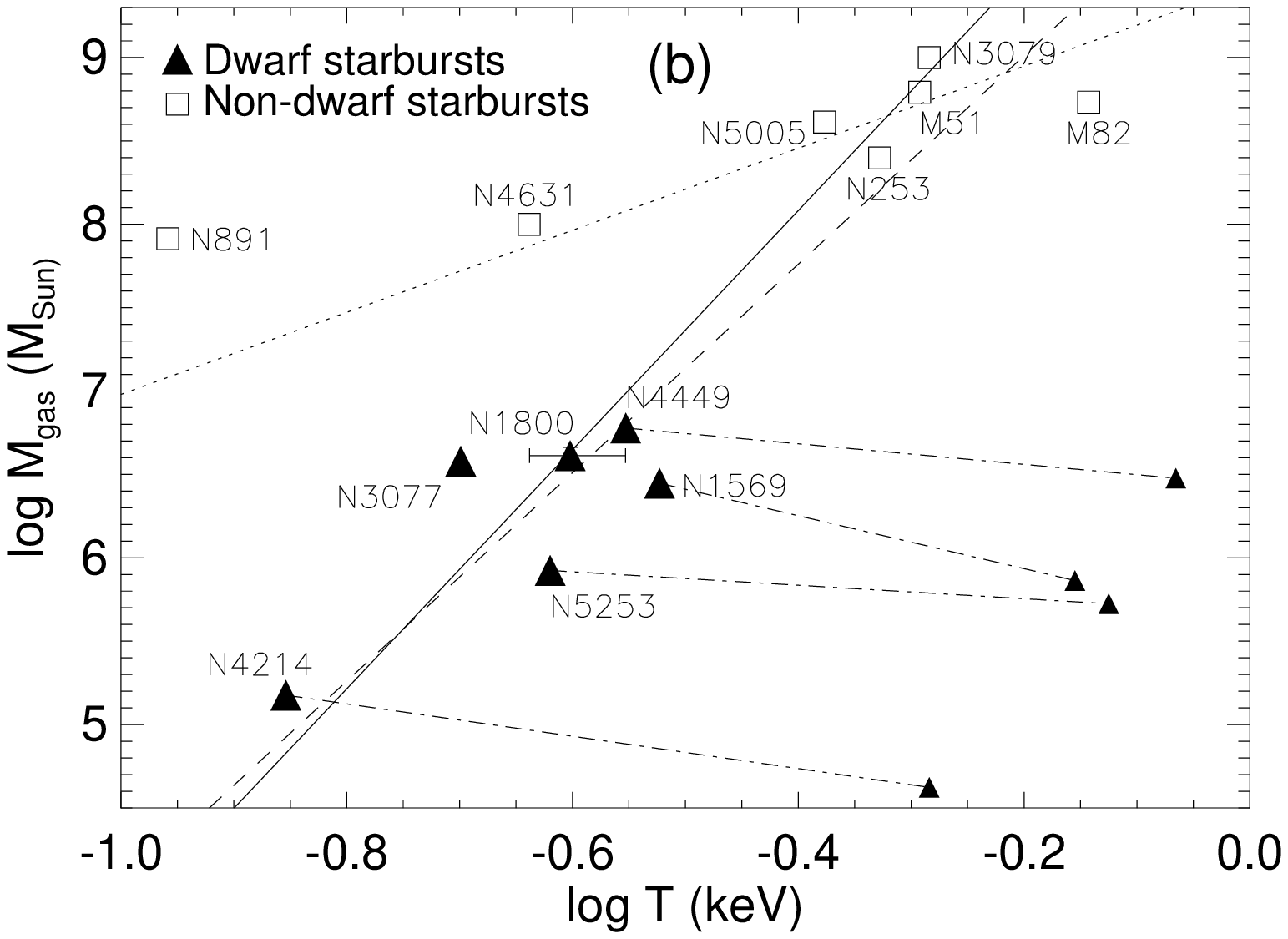}}
\mbox{\hspace{-0.5cm}
\epsfxsize=9.2cm
\epsfysize=6.8cm
\epsfbox{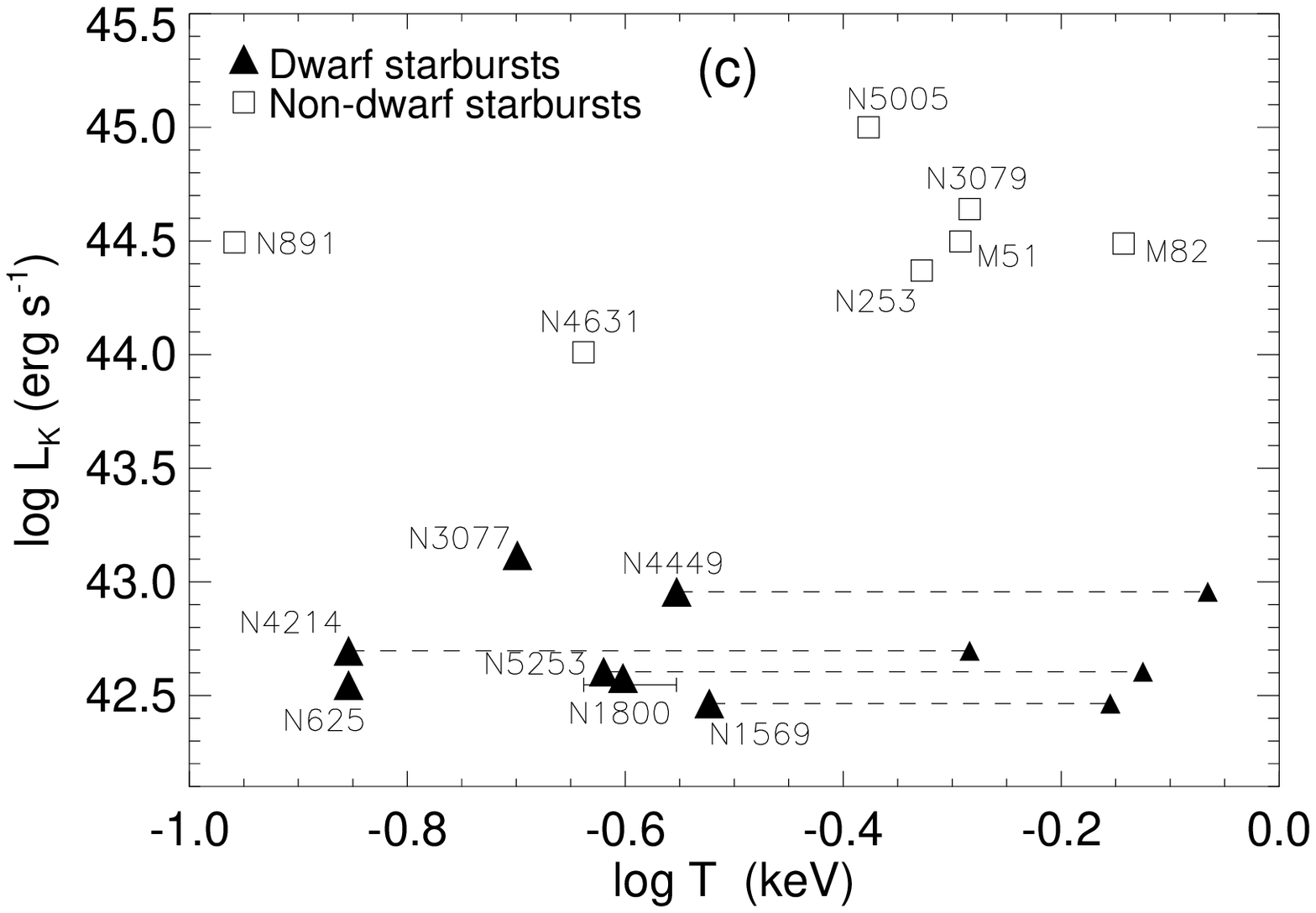}}
\caption{Temperature of the diffuse gas along with its
X-ray luminosity $L_X$ (a), mass $M_{gas}\eta^{1/2}$ (b), 
and the galactic 'mass' $L_K$ (c).
Where values for a second thermal component have been published,
this component is shown with smaller symbols, connected to the colder 
component by dash-dotted lines. Lines have the same meaning as for 
Fig.~\ref{fig,lxlfir}a, except that NGC\,891 has been omitted from the
regression analyses of the non-dwarf and full samples in 
Fig.~\ref{fig,lxlfir4}a and \ref{fig,lxlfir4}b, whereas NGC\,3077 {\em is}
included in the dwarf and full samples of Fig.~\ref{fig,lxlfir4}b. Errors on
$M_{gas}$ for NGC1800 are based on the fractional errors on the 
spectral normalization.}
\label{fig,lxlfir4}
\end{center}
\end{figure}
The situation is complicated slightly by the
presence of a second thermal component in some systems. Here the two 
components are plotted separately where possible. Fig.~\ref{fig,lxlfir4}a 
shows $L_X$ versus $T$, in which, again, NGC\,1800 resumes a fairly typical
position. When excluding NGC\,3077 from the dwarf sample, there is marginal
evidence ($t_s=1.2$) for $L_X$ correlating with $T$ for 
these galaxies. There is a much stronger trend for the normal starburst
subsample ($t_s=5.7$ with NGC\,891 excluded, see below), which is also
reflected in the high statistical significance of the full-sample relation 
($t_s=5.4$, again excluding NGC\,891).
For the hotter of the two thermal components there is again a suggestive 
trend ($t_s=1.2$), but any comparison to the behaviour of the cold component
in dwarfs or to normal starbursts is hampered by the small numbers involved. 
While this restriction also inhibits any firm conclusions for the cold
component, the position of the best-fitting relation for the RP01 starbursts 
does seem to indicate that dwarf bursts are slightly hotter than would be 
suggested by an extrapolation of this relation to lower X-ray luminosities. 
However, also evident from Fig.~\ref{fig,lxlfir4}a is
that in systems where a second, hotter thermal component has been detected,
the colder component dominates the soft X-ray output (and hot gas mass, as
discussed below). This probably explains why systems in which only one 
component is detected tend to have $T$ consistent with that of the colder 
component in two-component systems. It could be speculated that more
sensitive observations would reveal evidence for a hot secondary component in
most, if not all, systems, including the normal starbursts in 
Fig.~\ref{fig,lxlfir4}a observed by {\em ROSAT}.
If so, the indication that dwarfs are hotter than expected from normal
starbursts could be an artefact of splitting $L_X$ and $T$ of the dwarfs 
into two components. But, as can be judged from Fig~\ref{fig,lxlfir4}a, even
if adding $L_X$ of the hot component to that of the cold component while 
assuming $T\equiv T_{cold}$, all dwarfs (exempting NGC\,3077) still lie at or 
to the right of the normal starburst regression line shown in the figure.

In Fig.~\ref{fig,lxlfir4}b, $T$ is correspondingly plotted along with the 
mass $M_{gas}\eta^{1/2}$ of X-ray emitting gas. Again, where a second thermal
component has been 
detected, values published for both components are plotted. NGC\,625 is not
included, because it lacks a published value for $M_{gas}$. Aside from the 
wind geometry assumed in the published analyses 
(an ellipsoidal volume for NGC\,1800, cylindrical ones for NGC\,1569 and 
NGC\,4214, and a spherical geometry for the remainder), 
the plotted gas masses $M_{gas}$ have all been consistently derived 
as outlined in Section~\ref{sec,hot}. 
Hence, Fig.~\ref{fig,lxlfir4}b should provide a fair 
comparison between different galaxies assuming that the filling factor 
$\eta$ does not vary substantially between them. Note, however, that the
value for NGC\,3077 as derived by \citet{ott2003} assumes $Z$ = Z$_\odot$. 
With resulting $M_{gas}$ scaling as $Z^{-0.5}$, the plotted value for this 
galaxy is thus a lower limit for any subsolar metallicity (but note also that
the ionized ISM of NGC\,3077 has $Z$ close to solar; \citealt{calz2004}).
Though dealing with small numbers, there seems to be a weak tendency for 
dwarfs with larger gas mass to display hotter gas, and this applies to both 
the 'cold' ($t_s=1.3$) and 'hot' ($t_s=1.9$) thermal components
(RP01 find a more significant trend, $t_s=3.9$, for their starbursts). 
The correlation does not seem
to arise from a scaling with total galaxy mass for either component, as can
be judged from Fig.~\ref{fig,lxlfir4}c (for the cold
component, $t_s=-0.5$ for dwarfs and 0.4 for normal starbursts, and $t_s=0.6$
for the hot component in dwarfs), although the combined sample shows a 
marginally significant trend ($t_s=1.8$, which however reduces to $t_s=1.0$
if NGC\,891 is removed from the sample). Again, dwarf winds appear to be 
hotter for a given gas mass than expected on the basis of starbursts in more 
massive galaxies, an indication which, once more, cannot be explained simply 
by our splitting of the X-ray gas in some dwarfs into two components.

Mainly due to its low gas temperature, NGC\,891 is seen to fall well outside 
the ($L_X,T$) and ($M_{gas},T$) regions occupied by other starburst galaxies.
Although the galaxy is actively forming stars 
at present and further shows evidence for extraplanar diffuse X-ray emission 
which is possibly related to star formation processes in the disc, it remains
unclear whether this emission represents an outflowing 
galactic wind (see the discussion in \citealt{stri2004}, who furthermore 
derive a mean temperature of $T=0.23$ keV for the off-disc gas, as opposed to 
$T=0.11\pm 0.03$ keV found by RP01 for the diffuse emission). 
Moreover, the FIR 'temperature' $S_{60}/S_{100}$ of the galaxy is only 
0.31, placing it in the 'normal' (quiescent) rather than 'starburst' 
subsamples of \citet{kenn1998} and \citet{stri2004}. Finally, the galaxy 
shows evidence for a hot halo extending to $\sim 5$ kpc above the disc. 
The gas temperature listed by RP01 for this galaxy is likely to reflect 
the properties of this corona-like gas to a significant degree 
(see also \citealt{read1997}). Given these circumstances, it seems sensible 
to exclude NGC\,891 from regression analyses involving $T$ of the X-ray gas in
starbursts. NGC\,4490 is also excluded in this context, as no value of
$T$ has been published for this galaxy.

\begin{table*}
\centering
\begin{minipage}{164mm}
\caption{\protect{Selected results of regression analyses assuming relations 
of the form $Y = mX+c$, with $X$ and $Y$ unweighted. 
Errors on the regression coefficients $m$ and $c$ are $1\sigma$. 
Column 3 specifies the 
subsample used (D = dwarfs, N = normal starbursts from Read \& Ponman 2001, 
All = D + N) and the number [$N$] of galaxies in the subsample. 
Column 6 gives the Spearman rank-order correlation coefficient $r_s$,
and column 7 its significance $t_s$ (see text for details). 
Column 8 specifies the figure depicting the derived relation. $L_X$ is 
assumed to be in the 0.1--2 keV band.}}
\label{tab,regress}
\begin{tabular}{cccrrrrc} \hline 
\multicolumn{1}{c}{$Y$} &
\multicolumn{1}{c}{$X$} &
\multicolumn{1}{c}{Galaxy sample [$N$]} &  
\multicolumn{1}{c}{$m$} &  
\multicolumn{1}{c}{$c$} & 
\multicolumn{1}{c}{$r_s$} & 
\multicolumn{1}{c}{$t_s$} &
\multicolumn{1}{c}{Fig.} \\ \hline 
log $L_X$ & log $L_{FIR}$ & D [7]               & $2.01\pm1.48$ & $-45.45\pm0.60$ & 0.50 & 1.3 & -- \\ 
log $L_X$ & log $L_{FIR}$ & D, no N3077 [6]     & $1.21\pm0.38$ & $-12.29\pm0.24$ & 0.77 &	2.4  & \ref{fig,lxlfir}a \\
log $L_X$ & log $L_{FIR}$ & All [15]	        & $0.82\pm0.09$ & $4.41\pm 0.20$ & 0.93 & 9.0  & -- \\ 
log $L_X$ & log $L_{FIR}$ & All, no N3077 [14]  & $0.88\pm0.07$ & $1.75\pm0.13$ & 0.96	& 11.3 & \ref{fig,lxlfir}a \\
 \\
log $L_X$ & log $L_B$ & D [7]                   & $2.88\pm0.59$ & $-83.67\pm0.14$& 0.89	&  4.4 & \ref{fig,lxlfir}b \\
log $L_X$ & log $L_B$ & D, no N3077 [6]	        & $2.36\pm0.54$ & $-61.50\pm0.14$& 0.89	& 3.8  & -- \\  
log $L_X$ & log $L_B$ & All [15]		& $1.15\pm0.15$	& $-10.09\pm0.15$ & 0.85	& 5.9 & -- \\
log $L_X$ & log $L_B$ & All, no M82 [14]	& $1.05\pm0.11$	& $-5.99\pm0.15$ & 0.94	& 9.8 & \ref{fig,lxlfir}b \\	
 \\
log $L_{FIR}/L_K$ & log $L_K$ &  N [8]		& $-0.53\pm0.23$& $22.87\pm0.15$ & $-0.40$& $-1.1$ & -- \\
 \\
log $L_X/L_B$	& log $L_{FIR}/L_B$ & D, no N3077 [6]	& $1.49\pm0.96$ & $-3.05\pm0.34$ & 0.66	 & 1.7 & \ref{fig,lxlfir3} \\
log $L_X/L_B$	& log $L_{FIR}/L_B$ & All [15]  & $0.89\pm0.26$ & $-3.42\pm0.23$ & 0.29	& 1.1 & \ref{fig,lxlfir3} \\
log $L_X/L_B$	& log $L_{FIR}/L_B$ & All, no N3077 [14]& $0.95\pm0.17$ & $-3.48\pm0.14$ & 0.52 & 2.1 & \ref{fig,lxlfir3} \\
 \\
log $L_X/L_K$	& log $L_{FIR}/L_K$ & D, no N3077 [6]	& $1.44\pm0.47$ & $-3.02\pm0.28$ & 0.77	& 2.4 & -- \\
log $L_X/L_K$	& log $L_{FIR}/L_K$ & N [8]		& $1.11\pm0.16$ & $-3.51\pm0.13$ & 0.69	& 2.3 & -- \\
log $L_X/L_K$	& log $L_{FIR}/L_K$ & All, no N3077 [14]& $1.51\pm0.36$ & $-3.14\pm0.20$ & 0.71	& 3.4 & -- \\
 \\	
log $L_{X,cold}$ & log $T_{cold}$ & N, no N4490/N891 [6] & $2.71\pm0.48$ & $41.14\pm0.20$ & 0.94 & 5.7 & \ref{fig,lxlfir4}a \\
log $L_{X,cold}$ & log $T_{cold}$ & All, no N4490/N891 [13] & $4.50\pm0.60$ & $41.76\pm0.28$ & 0.85 & 5.4 & \ref{fig,lxlfir4}a \\
 \\
log $L_{X,hot}$ & log $T_{hot}$ & D [4]		& $3.25\pm0.50$ & $38.80\pm0.01$ & 0.63	& 1.2 & -- \\
 \\
log $M_{gas,cold}$ &	log $T_{cold}$ & D, no N625 [6] & $6.25\pm1.37$	& $10.26\pm0.31$ & 0.54	& 1.3 & \ref{fig,lxlfir4}b \\
log $M_{gas,cold}$ &	log $T_{cold}$ & N, no N4490/N891 [6] & $2.46\pm0.65$ &	$9.44\pm0.18$ & 0.77 & 2.4 & \ref{fig,lxlfir4}b \\
log $M_{gas,cold}$ &	log $T_{cold}$ & All, no N891/N4490/N625 [12] & $7.17\pm0.92$ & $10.95\pm0.33$ & 0.83 & 4.7 & \ref{fig,lxlfir4}b \\
 \\
log $M_{gas,hot}$ &	log $T_{hot}$ & D [4]	& $8.55\pm0.25$ & $7.02\pm0.06$  & 0.80 & 1.9 & -- \\
\hline \\
\end{tabular}
\end{minipage}
\end{table*}

\subsection{Dwarf vs normal starbursts}
An overall conclusion of the above comparisons, which is
identical to that arrived at by RP01 for non-dwarf starbursts, is
that the diffuse X-ray emission in dwarf starbursts seems to be related to
star formation activity rather than galactic (stellar) mass. 
This contrasts with the
results for quiescent spirals but is in line with properties of more massive
starbursts. While there is an indication (at the $\sim 2\sigma$ level) that
the diffuse $L_X$ of dwarf starbursts has a steeper dependence on $L_B$ or 
$L_K$ than in 'normal' starbursts, dwarf starbursts as a class appear to 
conform to the scaling relations obeyed by normal starburst galaxies.
Dwarf bursts can therefore, to a large degree, be considered 
down-scaled versions of normal starburst galaxies. 
Whether the properties of NGC\,1800 are affected by being in a group or not, 
is not easily addressed by these comparisons, given the small size of our
dwarf starburst sample. Specifically, it
could be imagined that the wind X-ray luminosity of NGC\,1800 could be 
enhanced due to compression by the ambient group ICM, but there is no such 
hint to be taken from Figs.~\ref{fig,lxlfir}--\ref{fig,lxlfir3}.

The only notable difference between dwarf and normal starbursts emerging 
from these results is that the winds of dwarfs appear to be somewhat hotter 
than expected from an extrapolation of normal starbursts to lower $L_X$ and,
particularly, $M_{gas}$ 
(Figs.~\ref{fig,lxlfir4}a and \ref{fig,lxlfir4}b). 
One possible explanation is that 
there is a lower limit to the temperature generated in superbubbles,
perhaps at $T\approx 0.10$--0.15 keV, which is
independent of galactic environment. An alternative is that {\em Chandra} is 
not sensitive to cooler gas, so that a powerful selection effect is at work 
on the derived gas temperatures. To investigate this, 
we performed a {\sc pimms} calculation, assuming a $Z=0.2$Z$_\odot$ plasma 
subjected to an absorbing column density $N_H=3\times 10^{20}$ cm$^{-2}$. 
For a fixed, unabsorbed 0.3--2 keV flux arriving at the ACIS-S detectors, 
$T\approx 0.6$ keV would maximize the observed 0.3--2 keV count rate. 
Compared to this, 80 per cent of the incoming photons remain undetected for 
$T=0.1$ keV (with 50 per cent at $T=0.2$ keV and 65 per cent at $T=0.15$ keV).
To safely detect low-$T$ components in starburst winds 
(characteristic temperatures $\la 0.10$ keV) in typical {\em Chandra} 
observations, these components would therefore have to be intrinsically
brighter than the entire NGC\,1800 wind by an 
order of magnitude. In the present X-ray dwarf sample, only the wind of the
unusual NGC\,3077 has the potential to meet this requirement.
This suggests that superbubble gas at such low temperatures would remain
undetected, and therefore that the apparent differences between dwarf and 
normal starbursts may be a result of observational selection effects. 
Support for this view comes from the different simulations of 
\citet{stri2000b}, in which very low-$T$ gas is present 
(down to $T\sim 10^5$ K), but the 
corresponding {\em characteristic} emission-weighted temperature as 
it would be inferred from {\em ROSAT} PSPC data is always $\ga 0.10$ keV.
Any clear detection of very low temperature gas in starburst winds would 
help resolve the issue.

\section{Summary}\label{sec,summary}

Diffuse X-ray emission has been unambiguously detected in NGC\,1800, 
the most distant dwarf
starburst galaxy for which this has yet been reported. The diffuse emission,
with a 0.3--3 keV luminosity of $1.3\pm 0.3\times 10^{38}$ erg s$^{-1}$,
accounts for at least 60 per cent of the total soft X-ray output of the
galaxy, with the remainder being contributed by five resolved point sources.
Two of these sources are located at the optical outskirts of the galaxy 
and may be unassociated with NGC\,1800 (possibly being background AGN), 
whereas the 
remaining three have X-ray luminosities consistent with X-ray binaries.
The morphology of diffuse emission suggests an elongated outflow of hot 
($T\simeq 0.25$ keV) gas,
similar to the winds seen in more nearby starburst galaxies, and
reaching distances of $\sim 2$ kpc above the galactic plane.
The northern tip of the diffuse X-ray emission coincides with an extended
H$\alpha$ structure. We suggest a scenario in which photoionization from
massive stars in the disc combines with shock ionization due to the expanding
wind, in producing and maintaining the H$\alpha$ emission of this structure.
There is no evidence for a hot, gaseous halo surrounding the galaxy, implying
an upper limit to the X-ray luminosity of such a halo in agreement with
expectations from simulations of disc galaxy formation.

NGC\,1800 is embedded in a small galaxy group, but we are unable to detect any
X-ray emission from intragroup gas, deriving an upper limit to the group 
X-ray luminosity of $L_X < 10^{41}$ erg s$^{-1}$. There is no clear evidence 
that the outflowing wind of NGC\,1800 is currently interacting with
any hot intracluster medium
in the group. Mechanical considerations suggest that the wind 
will probably be able to escape the extended H{\sc i} halo of the galaxy, 
particularly if the halo has a patchy distribution. Models including external 
pressure from intragroup gas suggest that the observable component of the
hot wind will not become confined by this gas either, thus enabling
the injection of wind energy and newly synthesized metals into 
the group ICM.

A first comparative study of the diffuse X-ray emission from dwarf 
starbursts as a class has been carried out. While lying at the low end of
the X-ray luminosity range spanned by dwarf starbursts with detected X-ray
emission, NGC\,1800 seems to be fairly 
typical of the small population studied in X-rays so far. We find that 
diffuse X-ray emission in 
these systems is related to starburst activity rather than galaxy mass,
equivalent to results inferred for more massive starburst galaxies. 
Specifically, there is
a clear correlation between the mass-normalized X-ray and far-infrared
luminosities, whereas the mass-normalized X-ray luminosity is largely
independent of galaxy mass. In this and other respects, the investigated 
properties of dwarf starbursts seem roughly consistent with 
an extrapolation to lower masses of starbursts in non-dwarf galaxies.
The X-ray luminosity and mass of the hot X-ray emitting gas in dwarfs 
appear to scale
weakly with emission-weighted mean temperature of this gas, a correlation 
which does not to seem to arise from a scaling of gas temperature with total 
galaxy mass. 
This indicates that the hot gas is far from being in hydrostatic equilibrium 
in such systems, consistent with the observed morphology of the gas.
In some systems, two thermal components are seen, but always with
the colder of the two components dominating the hot gas mass and diffuse 
X-ray emission.
Interestingly, there is a clear indication that the X-ray gas in dwarf
starbursts is generally hotter than would be expected from an extrapolation
of normal starbursts to low $L_X$ and $M_{gas}$. However, it cannot be ruled 
out at 
present that such a result simply reflects our limited ability to detect
very low-temperature gas in starburst winds.

\section*{Acknowledgments}
We thank the referee for comments which improved the clarity of this paper.
This work has made use of the 2~Micron All Sky Survey database, and the
Lyon-Meudon (LEDA) and NASA/IPAC (NED) extragalactic databases.
JR acknowledges support by the Danish Natural Science Research Council (SNF)
and by the Instrument Center for Danish Astrophysics (IDA).
TJP acknowledges the support of a Senior Fellowship from the Particle Physics
and Astronomy Research Council.\\

\label{lastpage}

\end {document}